\documentclass[a4paper, 10pt, leqno]{amsart}

\usepackage{amsmath}
\usepackage{graphicx}
\usepackage{rotating}
\usepackage{caption}
\usepackage{subcaption}
\usepackage{fancyhdr}
\usepackage{hyperref}
\usepackage{multirow,booktabs}
\usepackage{amsmath,amsthm,amsfonts,amssymb,amscd}

\usepackage{float}
\usepackage{epsfig}
\usepackage[utf8]{inputenc}
\usepackage[T1]{fontenc}

\usepackage[autolanguage]{numprint}

\usepackage{xcolor}
\usepackage{mdwlist}
\usepackage{amsaddr}

\usepackage[linesnumbered,ruled,vlined]{algorithm2e}
\usepackage[compatible]{algpseudocode}  

\usepackage[nameinlink,noabbrev,capitalise]{cleveref}
\usepackage{todonotes}

\newtheorem{theo}{Theorem}[section]
\newtheorem{lemme}[theo]{Lemma}
\newtheorem{cor}[theo]{Corollary}

\newtheorem{prop}[theo]{Proposition}
\newtheorem{remarque}[theo]{Remark}

\newtheorem{example}[theo]{Example}

\newcommand{\beq}{\begin{eqnarray}}
\newcommand{\enq}{\end{eqnarray}}
\newcommand{\be}{\begin{eqnarray*}}
\newcommand{\en}{\end{eqnarray*}}
\newcommand{\ben}{\begin{eqnarray*}}
\newcommand{\enn}{\end{eqnarray*}}

\newcommand{\Rd}{\mathbb R^d}
\newcommand{\R}{\mathbb R}
\newcommand{\N}{\mathbb N}

\newcommand{\dt}{\partial_t}

\newcommand{\demi}{\frac{1}{2}}

\newcommand{\dx}{\partial_x}
\newcommand{\dxx}{\partial_{xx}}
\newcommand{\dy}{\partial_y}

\newcommand{\E}{\mathbb E}

\newcommand{\lb}{\langle}
\newcommand{\rb}{\rangle}

\newcommand{\bi}{\begin{itemize}}
\newcommand{\ei}{\end{itemize}}

\newcommand{\cP}{\mathcal{P}}

\newcommand{\rmd}{\mathrm{d}}

\DeclareMathOperator{\supp}{supp}
\DeclareMathOperator{\co}{co}
\DeclareMathOperator{\interior}{int}
\newcommand{\Cpl}{\text{Cpl}}
\newcommand{\MCov}{{\rm MCov}}
\newcommand{\Ecal}{\mathcal{E}}
\newcommand{\Pcal}{\mathcal{P}}
\newcommand{\Wcal}{\mathcal{W}}
\newcommand{\Ncal}{\mathcal{N}}
\newcommand{\FL}{\star}
\newcommand{\HS}{\text{HS}}
\newcommand{\tr}{\mathrm{tr}}

\usepackage{bbm}
\usepackage{enumitem}

\newcounter{aspt}
\newcommand{\aspttag}{}

\crefname{aspt}{}{}
\Crefname{aspt}{}{}

\newcommand{\asptitem}[2]{
  \renewcommand{\aspttag}{#2}%
  \refstepcounter{aspt}%
  \item[{\rm (#1)}]%
}

\usepackage[backend=biber,natbib=true,bibencoding=utf8,style=authoryear,url=false,
  doi=false]{biblatex}
\begin{document}
\pagestyle{plain}

\title{The Martingale Sinkhorn Algorithm}
\author{Manuel Hasenbichler$^1$, Benjamin Joseph$^{2*}$, Gr\'{e}goire Loeper$^3$, Jan Ob\l{}\'{o}j$^{4\dag}$ and Gudmund Pammer$^5$}
\thanks{*This research has been supported by BNP Paribas Global Markets and the EPSRC Centre for Doctoral Training in Mathematics of Random Systems: Analysis, Modelling and Simulation (EP/S023925/1).}
\thanks{$\dag$The author would like to thank the Isaac Newton Institute for Mathematical Sciences, Cambridge, for support and hospitality during the programme \emph{Bridging Stochastic Control And Reinforcement Learning} where part of the work on this paper was undertaken. This work was partially supported by EPSRC grant no EP/R014604/1 and by a grant from the Simons Foundation.}
\address{
$^1$Institute of Statistics, Graz University of Technology\\
\texttt{manuel.hasenbichler@tugraz.at}\smallskip\\
$^2$Mathematical Institute and Christ Church, University of Oxford\\
\texttt{benjamin.joseph@maths.ox.ac.uk}\smallskip\\
$^3$BNP Paribas Global Markets\\ \texttt{gregoire.loeper@bnpparibas.com}\smallskip\\
$^4$Mathematical Institute and St John's College, University of Oxford\\ \texttt{jan.obloj@maths.ox.ac.uk} \\
$^5$Institute of Statistics, Graz University of Technology\\
\texttt{gudmund.pammer@tugraz.at}\smallskip\\
}


\begin{abstract}
We develop a numerical method for the martingale analogue of the Benamou--Brenier optimal transport problem, which seeks a martingale interpolating two prescribed marginals which is closest to the Brownian motion. Recent contributions have established existence of the optimal martingale under finite second moment assumptions on the marginals, but numerical methods exist only in the one-dimensional setting. We introduce an iterative scheme, a martingale analogue of the celebrated Sinkhorn algorithm, and prove that it yields a Bass potential in arbitrary dimension under minimal assumptions. In particular, we show that this holds when the marginals have finite moments of order $p > 1$, thereby extending the known theory beyond the finite-second-moment regime. The proof relies on a strict descent property for the dual value of the martingale Benamou--Brenier problem. While the descent property admits a direct verification in the case of compactly supported marginals, obtaining uniform control on the iterates without assuming compact support is substantially more delicate and constitutes the main technical challenge.
\end{abstract}
\maketitle

\section{Introduction} \label{sec:introduction}
This work contributes to the fields of optimal transport (OT), stochastic analysis and numerical analysis. Optimal transport, going back to \cite{Monge} and \cite{kanto1}, has proven to be a very rich and impactful field of mathematics. Its most classical problem, that of minimisation of quadratic cost among couplings of two marginal distributions, has motivated a string of seminal works including results of \cite{Br1, BB1, McGa}, and its entropic regularization links to the 
Schr\"odinger bridge problem studied in physics and led to powerful numerical algorithms, widely used across the machine learning community, see \cite{cuturi2013sinkhorn,cuturipeyre}. More recently, researches have studied OT problems with an additional martingale constraint. Originally motivated by the Skorokhod embedding problem, see \cite{genealogia,BeiglbockCoxHuesmann}, and applications in mathematical finance \cite{mass_transport,HenryLabordere:2014hta}, these problems have quickly opened up many beautiful mathematical challenges and led to further developments, including the so-called causal, or adapted, OT problems, see \cite{backhoff2020all}. 

In this paper, we focus on the martingale Benamou-Brenier (mBB) problem. \cite{BB1}, in their seminal work, introduced a fluid dynamic reformulation of the classical OT problem, which considers particles with a given initial and terminal distribution, moving in a vector field and minimising their distance to a constant velocity particle. 
Their seminal work had a profound impact. It provided an explicit interpretation of Wasserstein geodesics as flows of mass that minimize kinetic energy, giving an Eulerian interpretation for the Wasserstein geometry described in \cite{McGa}, matching and complementing the gradient flow viewpoint of \cite{JorKinOtt}. It opened the doors for new, PDE-driven, computational methods for OT problems, it influenced dynamic viewpoints on Entropic OT, see \cite{Leonard}, and modern application in the machine learning and stastical optimal transport, see \cite{cuturipeyre,ChewiNilesWeedRigollet2025}. 

The martingale version of the seminal Benamou--Brenier perspective considers martingales with prescribed initial and terminal distributions and minimises their distance to Brownian motion. The problem was introduced and studied by \cite{BeiglbockMBB} and its solution is given by the so-called Bass martingale, going back to \cite{bass1983skorokhod} and his solution to the Skorokhod embedding problem. It is also known as the \emph{stretched Brownian motion (sBM)}, and has recently been studied by \cite{backhoffveraguas2025existence,backhoff2023bass}. In analogy to the OT setting, the dynamic problem has its static counterpart, which is a weak martingale OT problem, in the spirit of \cite{gozlan2017kantorovich}. Like in the OT, both the dynamic and the static problem admit dual formulations, developed by \cite{HuesmannTrevisan} and \cite{backhoffveraguas2025existence} respectively. 

Despite significant theoretical advances, numerical methods for the mBB problem in $\R^d$, for $d \geq 2$, have not been developed.
This paper solves this important challenge. The one-dimensional case was considered already by \cite{conze2021bass}, who relied on the explicit formula for the OT one-dimensional map to develop a fixed-point-like iteration scheme. Its convergence was established separately in a recent study by \cite{acciaio2025calibration}, but the proof relied on the particular features of the one-dimensional problem and, as we will see, does not generalize to higher dimensions, which required genuinely novel ideas. Our methods rely on both the static and dynamic formulations, which allows us to build intuition for our results both from a PDE perspective, as well as convex duality perspective. Our iteration scheme can be seen as a martingale analogue of the celebrated Sinkhorn algorithm, or the iterative proportional fitting procedure proposed by \cite{cuturi2013sinkhorn} to solve entropic OT using approach of \cite{Sinkhorn}. 

The paper is organised as follows. Sections \ref{sec:BB}-\ref{sec:mBB} review the classical Benamou–Brenier formulation and its martingale counterpart, and Section \ref{subsec:main_results} presents the main theorem and introduces the key ingredients used in its proof. Section \ref{sec:PDEperspective} places the duality in a dynamic context and provides a PDE derivation. Section \ref{sec:Sinkhorn} recalls the classical Sinkhorn algorithm and explains the terminology \emph{martingale Sinkhorn algorithm}. Further background and related work are collected in Section \ref{subsec:literature}. Section \ref{sec:convergence} contains the proof of the main result, including its most general formulation in Theorem \ref{thm:limit_points_MS}. Section \ref{sec:implementation_and_examples} discusses the numerical implementation of our scheme and presents examples. 
Auxiliary results appear in Appendix \ref{sec:appendix:auxiliary_results}.

\textbf{Notation.} We denote $\cP(\Rd)$ probability measures on $\Rd$, $\cP_p(\Rd)$ measures with finite $p^{\textrm{th}}$ moment and write $\operatorname{supp}(\mu)$ for the support of $\mu$. The pushforward of $\mu\in \cP(\Rd)$ by a map $T:\Rd\to \R^m$ is denoted $T_\# \mu = \mu \circ T^{-1}\in \cP(\R^m)$. Standard Gaussian distribution on $\Rd$ is denoted $\gamma$ and $*$ denotes the convolution operator. The Legendre--Fenchel (LF) transform of $\psi$ is given by $\psi^{\FL}(y):=\sup_{x\in\R^d}(y\cdot x - \psi(x))$. A convex function ${\psi\colon E \to \R\cup\{\pm\infty\}}$ is called \emph{proper} if $\psi(x) > -\infty$ for every $x \in E$ and its \emph{effective domain} ${\operatorname{dom}(\psi) := \{x\in E\colon \psi(x) < \infty\}}$ is non-empty. 
For $S \subset \R^d$, we denote $\operatorname{ri}(S)$ the relative interior of $S$, i.e., its interior within its affine hull $\operatorname{aff}(S)$, and $\co(S)$ the convex hull of $S$. For a function $f\colon \Rd \to \R$, we denote by $\partial f(M)$ the set of subgradients of $f$ at $M$. We will introduce additional notation below when and as required. For further convex-analysis conventions, we refer to \cite{R}.

\section{Benamou-Brenier perspective and the Martingale Sinkhorn Algorithm}
\subsection{The classical OT problem and its Benamou-Brenier formulation}\label{sec:BB}
Given $\mu,\nu\in \cP_2(\Rd)$, the quadratic cost OT problem in its static formulation, and its dynamic Benamou-Brenier counterpart, is given as
\begin{equation}\tag{OT}\label{eq:OT}
    \inf_{\pi \in \Cpl(\mu,\nu)} \int |x-y|^2 \,\pi(\rmd x,\rmd y) = \inf_{\substack{X_0\sim \mu, X_1\sim \nu\\ X_t=X_0+\int_0^t b_s \mathrm{d}s}} \E \left[\int_0^1|b_s |^2\,\mathrm{d}t\right], 
\end{equation}
where $\Cpl(\mu,\nu)\subset \cP_2(\Rd\times \Rd)$ denotes the set of couplings with marginals $\mu$ and $\nu$. Note that the optimiser $X^*$ is the same if we replace $|b_s|^2$ by $|b_s-b|^2$, for any constant $b$, hence the interpretation of keeping $X$ as close as possible to a constant velocity particle. \cite{Br1} in his seminal work showed that, under mild regularity assumptions on $\mu$, the optimal coupling $\pi^{\ast}$ is obtained with $\nu$ being the pushforward of $\mu$ via a gradient of a convex function $v$, i.e., $\pi^{\ast}=(Id, \nabla v)_\# \mu$. We have $(X^*_0,X^*_1)\sim \pi^{\ast}$ and the intermediate distributions $\mu_t \sim X^*_t$ trace the Wasserstein geodesic between $\mu$ and $\nu$, linking the fluid mechanics/PDE/variational approach with the geometric insights of \cite{MC1}. To illustrate these objects, 
expanding the square and noting that $\int |x|^2 + |y|^2 \rmd \pi$ is constant for $\pi\in \Cpl(\mu,\nu)$, it is convenient to rewrite the OT problem in an equivalent formulation as follows: 
\begin{equation}\label{eq:MCov}\tag{MCov}
\begin{split}
    \MCov(\mu,\nu) := &\sup_{\pi \in \Cpl(\mu,\nu)} \int x\cdot y \,\pi(\rmd x,\rmd y) = \int x \cdot \nabla v(x) \,\mu(\rmd x) = \int v \, \rmd \mu + \int v^{\FL} \, \rmd \nu\\
    =  &\min_{\psi \text{ l.s.c., convex}} \int \psi \,\mathrm{d}\mu + \int \psi^{\FL} \, \mathrm{d}\nu.
\end{split}
\end{equation}

The convex function $v$ is known as the Brenier--McCann potential between $\mu$ and $\nu$, characterised by $\nu=(\nabla v)_\#\mu$. The second equality above follows directly from the properties of the LF transform, $\psi(x)+\psi^{\FL}(\nabla \psi(x))=x\cdot \nabla \psi(x)$, and the last equality is the OT duality. 
Finally, under mild regularity assumptions, we see that $\mu=(\nabla v^{\FL})_\# \nu$ and, by symmetry, it follows that $v^{\FL}$ is the Brenier--McCann potential between $\nu$ and $\mu$.

\subsection{The martingale Benamou-Brenier problem} \label{sec:mBB}
Given $\mu_0,\mu_1\in \cP_2(\Rd)$, we say that $\pi$ is a martingale coupling of $\mu_0$ and $\mu_1$ if $\pi\in \Cpl(\mu_0,\mu_1)$ and 
$\E^\pi[Y \mid X] = X$ where $X(x,y) = x$ and $Y(x,y) = y$ denote the coordinate projections on $\mathbb{R}^d \times \mathbb{R}^d$. Equivalently,
$\int y \,\pi_x(\rmd y) = x$ $\mu_0$-a.e., where $\pi(\rmd x,\rmd y) = \mu_0(\rmd x)\pi_x(\rmd y)$ is the disintegration of $\pi$.
We denote by $\Cpl_M(\mu_0,\mu_1)$ the set of all such martingale couplings $\pi$ and note that by a theorem of \cite{strassen1965existence}, $\Cpl_M(\mu_0,\mu_1)$ is non-empty if and only if the marginals are in convex order, that is,
\[
    \mu_0 \leq_\mathrm{cvx} \mu_1 \quad \Longleftrightarrow\quad \int v \,\mathrm{d}\mu_0 \leq \int v \,\mathrm{d}\mu_1 \text{ for all convex } v \colon \R^d\to\R.
\]
Under this condition, one may consider the left-hand side of \eqref{eq:OT} with the infimum restricted to $\pi\in \Cpl_M(\mu_0,\mu_1)$. In this case, however, the problem becomes degenerate, since the value of the integral does not depend on the choice of $\pi$. Various alternative cost functions have been considered in the context of martingale optimal transport; see \cite{beiglbock2022bachelier,henry2017model} and the references therein. In the present paper we work with the right-hand side of \eqref{eq:OT}, which leads to the following martingale analogue introduced in \cite{BeiglbockMBB}:
\begin{align} \label{eq:mbb} \tag{mBB}
    \mathbf{MT}_{\mu_0,\mu_1}&=\inf_{\substack{M_0\sim \mu_0, M_1\sim \mu_1\\ M_t=M_0+\int_0^t \sigma_s \,\mathrm{d}B_s}} \E \left[\int_0^1|\sigma_t - \bar\sigma I_d|_{\HS}^2\,\mathrm{d}t\right],
\end{align}
where $|\cdot |_{\HS}$ denotes the Hilbert--Schmidt norm, $\bar \sigma >0$ and $I_d$ is the identity matrix in $\R^d$. The infimum is taken over all filtered probability spaces supporting a Brownian motion $B$, over progressively measurable $\Rd\times\Rd$-valued processes $\sigma$, and over martingales $M$ defined via It\^o integral representation. Note that any admissible process $M$ in~\eqref{eq:mbb} induces a \emph{martingale coupling}  $\pi := \mathrm{Law}(M_0, M_1)\in \Cpl_M(\mu_0,\mu_1)$. \eqref{eq:mbb} is called the \emph{martingale Benamou-Brenier} problem and \cite{BeiglbockMBB} showed it admits a unique (in distribution) optimiser, $\pi^{\rm SBM}$, which is referred to as \emph{stretched Brownian motion}.

In analogy with the OT case, expanding the square allows us to rewrite \eqref{eq:mbb} in the equivalent form
\begin{equation} \label{eq:mbb_sup}
    \mathbf{P}_{\mu_0,\mu_1} = \sup_{\substack{M_0\sim \mu_0, M_1\sim \mu_1\\ M_t=M_0+\int_0^t \sigma_s \,\mathrm{d}B_s}} \E \left[\int_0^1\tr\left(\sigma_t\right)\,\mathrm{d}t\right],
\end{equation}
in the sense that the two problems share the same optimiser and 
\begin{equation}\label{eq:2pbrelated}
\mathbf{MT}_{\mu_0,\mu_1} = \bar\sigma^2 d+\int |x|^2 \,(\mu_1(\rmd x)-\mu_0(\rmd x)) - 2 \bar\sigma\mathbf{P}_{\mu_0,\mu_1}. 
\end{equation}
The new problem, in turn, can be reformulated as a static weak optimal martingale transport which also admits a natural dual problem, as shown by \cite{backhoffveraguas2025existence}, 
\begin{equation} \label{eq:dmbb} \tag{dmBB}
    \mathbf{P}_{\mu_0,\mu_1} := \sup_{\pi \in \Cpl_M(\mu_0,\mu_1)} \int \MCov(\pi_x,\gamma)\, \mu_0(\rmd x) = 
    \inf_{\substack{\psi \in L^1(\mu_1),\\ \psi \text{ convex}}}  \left( \int \psi \,\mathrm{d}\mu_1 - \int \psi^C \,\mathrm{d}\mu_0 \right),    
\end{equation}
where $\gamma\sim (B_1-B_0)$ and $\psi^C:= (\psi^{\FL} * \gamma)^{\FL}$. We write 
\[\mathcal{E}(\psi) := \int \psi \,\mathrm{d}\mu_1 - \int \psi^C \,\mathrm{d}\mu_0\] for the dual objective.
Note that, while the derivation in \cite{backhoffveraguas2025existence} is carried out under second–moment assumptions, these are not necessary. 
Indeed, strong duality holds under finite $p$-moments for some $p>1$, together with the same structural conditions; see Theorem~\ref{app:thm:strong_duality_mBB}. 
We take these more general assumptions as the framework for our analysis below.

Throughout we assume that the pair $(\mu_0,\mu_1)$ is in convex order and \emph{irreducible}, meaning that for any Borel sets $A,B \subset\R^d$ with $\mu_0(A)>0$ and $\mu_1(B)>0$, there exists a martingale coupling $\pi \in \Cpl_M(\mu_0,\mu_1)$ such that $\pi(A\times B)>0$. This fundamental notion was first studied by Beiglb\"ock and Juillet~\cite{beiglboeck2016problem} on the real line, and later extended to $\mathbb{R}^d$ for any $d\geq1$ by \cite{de2019irreducible}, see also \cite{2017arXiv170208433O}. When irreducibility fails, the martingale OT problems split into separate subproblems supported on non-communicating regions of $\Rd$. Equivalently, the dual problem \eqref{eq:dmbb} is attained by a lower semi-continuous (l.s.c.) convex potential $v$ if and only if the pair $(\mu_0,\mu_1)$ is irreducible. In this case, the primal optimiser in \eqref{eq:mbb} is the \emph{Bass martingale} $(M_t)_{t \in [0,1]}$ defined by
\begin{equation}\label{eq:Bassmgdef}
    M_t := \E\!\left[\nabla v^{\FL}(B^\alpha_1) \mid B^\alpha_t\right], \qquad t \in [0,1],
\end{equation}
where $(B_t^\alpha)_{t \in [0,1]}$ is a Brownian motion with initial law $B_0^\alpha \sim \alpha := (\nabla v^C)_\# \mu_0$. Equivalently, $\pi^{\rm SBM}$ admits the static representation
\begin{equation}\label{eq:Basscoupling}
    \pi^{\rm SBM} = (\operatorname{proj}_x, T)_\# (\mu_0 \otimes \gamma), \qquad T(x,z) := \nabla v^{\FL}\!\left(z + \nabla v^C(x)\right).
\end{equation}
We refer to $v$ and $\alpha$ as the \emph{Bass potential} and the \emph{Bass measure}, respectively.

\subsection{Main Results} \label{subsec:main_results}
Our contribution is to propose a numerical algorithm for computing a Bass potential and to prove its convergence under very general assumptions. By analogy with the classical Sinkhorn algorithm---see \cref{sec:Sinkhorn} for a detailed discussion---we refer to the resulting method as the \emph{Martingale Sinkhorn Algorithm}.

\begin{algorithm}[ht]
\caption{The Martingale Sinkhorn Algorithm}\label{alg:MSinkhorn}
\KwIn{$\mu_0,\mu_1 \in \mathcal{P}_{p}(\R^d)$ for some $p > 1$, convex potential $v_0$\\}
\KwOut{$v_\texttt{max\char`_iter}$, $\alpha_\texttt{max\char`_iter}$} \vspace{1em}
\For{$i=1$ to \texttt{max\char`_iter}}{
	$\alpha_i \gets \bigl(\nabla v_{i-1}^C\bigr)_\# \mu_0$ \hfill ($v_{i-1}^C$ is the Brenier--McCann potential of $(\mu_0, \alpha_i)$) \hfill \\

    Find $v_i$ such that $\mu_1=(\nabla v^{\FL}_i)_\# (\alpha_i * \gamma)$ \hfill ($v^{\FL}_i$ is the Brenier--McCann potential of $(\alpha_i * \gamma, \mu_1)$)
}
\Return	$v_\texttt{max\char`_iter}$, $\alpha_\texttt{max\char`_iter}$
\end{algorithm}

We discuss the implementation of this algorithm in \cref{sec:implementation_and_examples}. Our main theoretical result asserts convergence of the above algorithm. We give here a shorter statement and refer to \cref{thm:limit_points_MS} for the full result. 

\begin{theo}[Limit points of the Martingale Sinkhorn algorithm] \label{thm:convergence_MSinkhorn_simple}
    Let $\mu_0,\mu_1 \in \mathcal{P}_{p}(\R^d)$ for some $p > 1$ be in convex order and irreducible, and let $v_0 \in L^1(\mu_1)$ be l.s.c.\ convex with $v_0^{\FL} * \gamma$ proper. Let $(v_i,\alpha_i)_{i\in\N}$ be the iterates generated by \cref{alg:MSinkhorn}. Then, after suitable affine normalisation of $(v_i)_{i\in\N}$, every accumulation point of $(v_i)_{i\in\N}$ on ${I_{\mu_1} := \operatorname{ri}\bigl(\operatorname{co}(\operatorname{supp}(\mu_{1}))\bigr)}$ is a Bass potential. Moreover, for any subsequence $(i_j)_{j\in\N}$ such that $v_{i_j}$ converges pointwise on $I_{\mu_1}$, the measures $(\alpha_{i_j})_{j\in\N}$ converge weakly in $\Pcal(\R^d)$ to the Bass measure associated with the limiting Bass potential.
\end{theo}
 
It follows by strong duality (see e.g.\ \cref{app:thm:strong_duality_mBB}) that if the Bass potential is $\mu_1$-integrable, the Martingale Sinkhorn Algorithm attains the dual problem~\eqref{eq:dmbb}. To our knowledge, this provides the first constructive existence proof of a Bass measure and associated potential in $\Rd$ under minimal assumptions. We also obtain existence of an optimal martingale in arbitrary dimension when the marginals admit finite moments of order $p>1$, thereby extending the known existence results for the martingale Benamou--Brenier problem, which required finite second moments of the marginals. 

In dimension one, we further show that the Bass potential is unique and attains~\eqref{eq:dmbb} (see \cref{prop:unique_basspotential_R1}). This implies that \cref{alg:MSinkhorn} converges. In this case, the algorithm can be re-written, with the two steps combined into one, as the fixed-point problem of \cite{conze2021bass}. Our work thus extends the convergence results of \cite{acciaio2025calibration}, who worked under strong regularity assumptions on $\mu_1$. Moreover, no a priori specification of the Bass measure is required, see \cref{subsec:literature} for more discussion. 

\begin{figure}[t]
    \centering
    \includegraphics[width=0.4\linewidth]{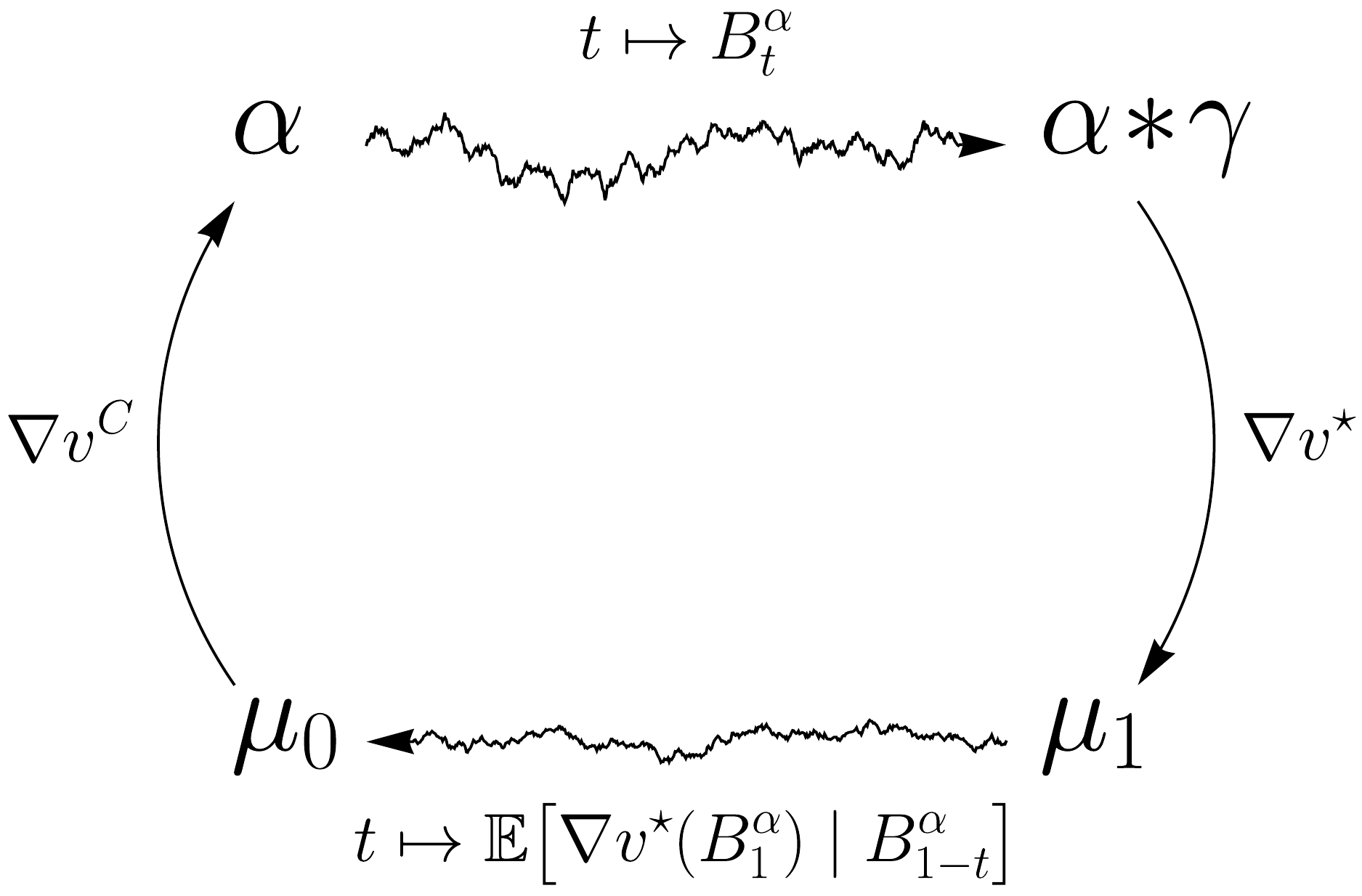}
    \caption{
        Schematic illustration of the relations characterising the Bass potential $v$ and Bass measure $\alpha$ that solve the Martingale Benamou--Brenier problem (see also \cref{sec:Sinkhorn}). Clockwise iteration of this diagram yields \cref{alg:MSinkhorn}, which decreases the dual objective function $\mathcal{E}$ after each full cycle.
    }
    \label{fig:MS_algorithm}
\end{figure}

The key mechanism behind the convergence of the Martingale Sinkhorn Algorithm is the strict descent of the dual objective in~\eqref{eq:dmbb}. When there exists a compact $K_0 \subset \mathbb{R}^d$ with $\supp \mu_0 \subseteq K_0 \subset \operatorname{ri}\bigl(\operatorname{co}(\supp \mu_{1})\bigr)$ (so that $(\alpha_i)_{i\in N}$ is compactly supported too; see \cref{lem:preservation_struc_props}), Brenier's theorem \eqref{eq:MCov} applies and clarifies the effect of the two updates in \cref{alg:MSinkhorn}:
First, the update $\alpha_i \gets (\nabla v_{i-1}^C)_\# \mu_0$ ensures that $v_{i-1}^C$ realises $\MCov(\mu_0,\alpha_i)$. Second, choosing $v_i$ such that $(\nabla v_i^\FL)_\# (\alpha_i * \gamma) = \mu_1$ makes $v_i^\FL$ the Brenier--McCann potential transporting $\alpha_i*\gamma$ to $\mu_1$. Unless $\alpha_i$ already is a Bass measure associated with $(\mu_0,\mu_1)$ and $v_i=v_{i-1}$, we obtain the inequalities
\begin{align*}
\int v_{i-1}^C \rmd \mu_0 + \int v_{i-1}^\FL * \gamma\,\rmd \alpha_i = \MCov(\mu_0,\alpha_i) &\leq  \int v_i^C \rmd \mu_0 + \int v_i^\FL * \gamma\,\rmd \alpha_i\\
 \int v_{i}\mathrm{d}\mu_1 + \int v_{i}^{\FL}*\gamma\,\mathrm{d}\alpha_i = \MCov(\alpha_{i} * \gamma,\mu_1) &< \int v_{i-1}^{\FL}*\gamma\,\mathrm{d}\alpha_i + \int v_{i-1}\,\mathrm{d}\mu_1.
\end{align*}

From these, it then readily follows that $\mathcal{E}(v_i) < \mathcal{E}(v_{i-1})$, as established in \cref{lem:strict_descent_compact}.

Importantly, if $\mu_0$ is not supported on a compact subset of $\operatorname{ri}(\operatorname{co}(\operatorname{supp} (\mu_{1})))$, this argument breaks down: As shown in \cref{ex:uniform_disk_to_circle} and \cref{rk:alphano1moment}, the iterates $(\alpha_i)_{i \in \N}$ generated by \cref{alg:MSinkhorn} need not have finite first moments, so that $\MCov(\mu_0,\alpha_i)=+\infty$. Moreover, the Bass potentials itself need not be $\mu_1$-integrable, as shown in \cref{ex:generic_bad_bass_potential}.
Hence, in \cref{sec:convergence} and in line with the approach of \cite{backhoffveraguas2025existence}, we work with a relaxed dual objective for which the strict descent discussed above still holds, see \cref{lem:strict_descent}.

A second, more subtle ingredient in the proofs is that boundedness of the dual objective values already ensures that, after an affine transformation, the potentials generated by the Martingale Sinkhorn Algorithm are uniformly locally bounded on $\operatorname{ri}\bigl(\operatorname{co}(\operatorname{supp}(\mu_1))\bigr)$, see \cref{lem:tightness}. This tightness relies only on the irreducibility of $(\mu_0,\mu_1)$ and notably does not require compact support or moment assumptions. It is this uniform control on the potentials that allows us to dispense with the compact-support assumption in the strict descent argument. Our full result is given in \cref{thm:limit_points_MS} below and establishes a stronger convergence result implying, in particular, that not only the potentials but also their gradients converge in a suitable sense.

We conclude with an example illustrating the subtlety of our arguments.
\begin{example} \label{ex:generic_bad_bass_potential}
    Let $(\mu_0,\mu_1)$ be uniform measures on $(-\tfrac{1}{2},\tfrac{1}{2})$ and $(-1,1)$ respectively, and define
    \[
        v(x) := \begin{cases}
            \displaystyle\frac{1}{1-x^2}, & x \in (-1,1),\\
            \infty, & \text{otherwise}.
        \end{cases}
    \]
    The potential $v$ is l.s.c.~and $2$-strongly convex, but not $\mu_1$-integrable. Thus $v^\FL$ is differentiable on $(-1,1)$, and $\nabla v^\FL$ is $\tfrac{1}{2}$-Lipschitz. In particular, $v^\FL * \gamma$ is strictly convex on $(-1,1)$ and has $\tfrac{1}{2}$-Lipschitz gradient as well. Hence $v^C := (v^\FL * \gamma)^\FL$ is differentiable and $2$-strongly convex, and one readily verifies that $\operatorname{dom}(v^C) = (-1,1)$. The associated Bass measure $\alpha := (\nabla v^C)_\# \mu_0 \in \Pcal(\R^d)$ admits finite $p$-moments only for $p < \tfrac{1}{2}$, since
    \[
        \int |y|^p \,\alpha * \gamma(\rmd y) = \int_{-1}^1 \left|\frac{2x}{(1-x^2)^2}\right|^p \frac{\rmd x}{2} = \infty \quad \text{if and only if } p \geq \tfrac{1}{2}.
    \]
    In particular, $\mu_0$ is not a Dirac measure, and therefore $\MCov(\mu_0,\alpha) = \infty$.
\end{example}

\medskip

\subsection{PDE perspective on duality and the Bass martingale}\label{sec:PDEperspective}
The \eqref{eq:mbb} problem is an instance of optimal transport under controlled semimartingale dynamics, studied by \cite{TanTouzi,HuesmannTrevisan,guo2021path}. Among other applications, these problems are used for non-parametric semimartingale projection to calibrate models in quantitative finance, see \cite{GuoLoWangLSV,guo2022joint}. For such solutions, the numerical methods rely on a PDE formulation of the dynamic dual problem, in a direct analogy to the motivation behind the work of \cite{BB1}. We discuss now briefly this perspective and show how it can be used to obtain the duality in \eqref{eq:dmbb}. Our aim is to provide additional intuition for the results and we keep the discussion formal and restrict ourselves to $d=1$. 

Consider a general Markovian formulation for the martingale Benamou--Brenier problem:
\begin{align}
\label{eq:mbbH}
\tag{mBB(H)} \mathbf{MT}_{\mu_0,\mu_1}^H&:=\inf_{\substack{M_0\sim \mu_0, M_1\sim \mu_1\\ M_t=M_0+\int_0^t \sigma(s,M_s) \,\mathrm{d}B_s}} \E \left[\int_0^1 H(\sigma(t,M_t)^2) \,\mathrm{d}t\right],
\end{align}
for a convex function $H$. Then, following \cite{HuesmannTrevisan,guo2021path}, we know that under suitable regularity assumption duality holds and the minimizer in \eqref{eq:mbbH} can be obtained by solving the dual problem 
\beq\label{eq:dual via PDE}
    \mathbf{MT}_{\mu_0,\mu_1}^H=\sup_\varphi\left\{\int_{\R}\varphi(1,x) \,\mu_1(\rmd x) -\int_\R \varphi(0,x)\,\mu_0(\rmd x)\right\},
\enq
where the supremum is taken over all super-solutions $\varphi$ of
\[
{\dt \varphi + H^{\FL}\left(\demi \dxx \varphi\right) \leq 0}.
\]
Further, if the supremum is attained by $\varphi$ then 
the optimiser is given by 
\begin{equation*}
 \sigma^2(t,x) =  \partial_y H^{\FL}\left(\frac{1}{2}\dxx\varphi(t,x)\right).
\end{equation*}
To recover \eqref{eq:mbb}, we consider the cost function 
\begin{equation*}
H(\beta) = \frac{1}{2}\left(\sqrt{\beta}-\bar\sigma\right)^2,\quad \textrm{and compute }
H^{\FL}\left(y\right) = \bar\sigma^2\frac{y}{1-2y},
\end{equation*}
and therefore the HJB equation becomes
\begin{equation*}
\dt \varphi + \frac{1}{2}\frac{\bar\sigma ^2 \dxx\varphi}{1-\dxx\varphi} = 0,
\end{equation*}
which is well studied in its log-normal form in \cite{LoMi1}, see also \cite{BoLoZo1, BoLoZo2, BoLoSonZou}.

Note that for $\varphi$ to solve the HJB equation, we have that $\dxx\varphi<1$. This means that the potential $v=\frac{x^2}{2}-\varphi$ is convex. It further satisfies
\begin{equation*}
\dt v + \demi \bar\sigma^2 \left(1-\frac{1}{\dxx v}\right)=0\quad \text{and}\quad \dt v^{\FL} + \demi \bar\sigma^2 (\partial_{yy} v^{\FL}-1)=0.
\end{equation*}
To obtain a representation aligned with \eqref{eq:dmbb}, take $\bar\sigma=1$ and observe that $v^\FL(t,y)-y^2/2$ solves the heat equation. In particular
\begin{align*}
 v^{\FL}(0,\cdot) &= \left(v^{\FL}(1,\cdot)-\frac{1}{2}\right)*\gamma,\quad \textrm{and hence} \quad v(0,\cdot) = \left(v^{\FL}(1,\cdot)*\gamma\right)^{\FL}+\frac{1 }{2}.
\end{align*}
We see that the condition that $\varphi$ solves the HJB equation is equivalent to the above relations between $v$, or $\varphi$, at times $0$ and $1$. Writing $\psi(x)=v(1,x)$ and recalling the relation \eqref{eq:2pbrelated}, we conclude that the duality \eqref{eq:mbbH}--\eqref{eq:dual via PDE} recovers the duality \eqref{eq:mbb_sup}--\eqref{eq:dmbb} in \cite{backhoffveraguas2025existence}.

Assuming the optimal potential $v$ exists, we can also easily recover the optimiser in \eqref{eq:mbb}. Indeed, note that the derivative of $v$, $\xi=\dx v$, solves the {\it linearized} equation
\begin{equation*}
\dt  \xi + \frac{\sigma^2}{2}\dxx\xi=0,\quad \text{with } \sigma=\frac{1}{\dxx v}=\frac{1}{1-\dxx\varphi}.
\end{equation*}
It follows that if we let $M^\sigma$ solve $\mathrm{d}M^\sigma_t = \sigma(t,M^\sigma_t)\mathrm{d}B_t$, then $Z_t:=\xi(t,M_t^\sigma)$ is a martingale. 
Moreover, we have
$$
\mathrm{d}\lb Z\rb_t=(\dx \xi(t,M_t^\sigma))^2 \sigma(t,M^\sigma_t)^2\mathrm{d}t =\mathrm{d}t,
$$
so, by L\'evy's theorem, $Z$ is a Brownian motion and we can write $M^\sigma_t = F(t, Z_t)$, where $F(t,y)=\dy v^{\FL}(t, y)$ solves the heat equation, $F(0,\cdot) = F(1,\cdot)*\gamma$. The densities $\mu(t,x)$ of the marginal distributions $M^\sigma_t$ satisfy the Fokker-Planck equation
$$\dt \mu  = \demi\partial_{xx} \left(\sigma^2 \mu\right)$$
which implicitly encodes the compatibility condition since both $\mu(0,\cdot) = \mu_0(\cdot)$ and $\mu(1,\cdot)=\mu_1(\cdot)$ are fixed. In particular, we recover the defining properties of the Bass martingale described earlier: 
$$ M^\sigma_t = F(t,Z_t) = \mathbb{E}\!\left[ F(Z_1) \mid Z_t\right], \text{ where}\quad
Z_0\sim \alpha =  \dx v(0,\cdot)_\# \mu_0\quad \textrm{and}\quad F(y)= F(1,y) = \dy v^{\FL}(1,y),
$$
noting that $v(0,\cdot) = v(1,\cdot)^C$. We now explore  these relations in more detail and compare them with those known for the Sinkhorn system.

\subsection{Analogy to the Schr\"odinger system and the Sinkhorn algorithm}\label{sec:Sinkhorn}
For notational convenience, we assume in this section that $\mu_0,\mu_1\in \cP_2(\Rd)$ admit densities, which we denote $\mu_0(x)$, $\mu_1(x)$. We also write $\gamma$ for both the Gaussian measure and its density. The classical Schr\"odinger problem, closely related to the Entropic OT, considers minimising the relative entropy, or the Kullback–Leibler (KL) divergence:
$$
\min_{\pi\in\Cpl(\mu,\mu_1)} \mathrm{KL}(\pi\,\|\,R)
\quad=\quad
\min_{\pi\in\Cpl(\mu,\mu_1)} \int_{\Rd\times\Rd}\!\log\!\Big(\frac{\rmd\pi}{\rmd R}\Big)\,\rmd\pi,
$$
where $R(\rmd x,\rmd y) := \gamma(x-y)\,\rmd x\,\rmd y$. There exist measurable functions $f,g:\Rd\to(0,\infty)$, such that the unique minimizer is $\pi^\star$ is given by
$$
\pi^\star(\rmd x, \rmd y)=f(x)\, \gamma(y-x) \,g(y)\,\rmd x\,\rmd y.
$$
The pair $(f,g)$ is unique up to a multiplication/division by a constant, and solves the \emph{Schr\"odinger system}
\begin{equation*}
\mu_0(x)=(g * \gamma)(x)\,f(x),\qquad
\mu_1(dy)=g(y)\,(f * \gamma )(y).
\end{equation*}
This system is approximated by the celebrated Sinkhorn algorithm, where each step is a Bregman \cite{bregman1966relaxation} projection, i.e., an iterative renormalization of the kernel $\gamma(x-y)$ to have the desired marginal:
\begin{align*}
\text{ Update } f \qquad \mu_0 =& (g^{i-1}*\gamma)f^{i},\\
\text{ Update } g \qquad \mu_1 =& g^{i}(f^{i}* \gamma).
\end{align*}
It is shown in \cite{Sinkhorn,cuturi2013sinkhorn} that this algorithm converges. To make the analogy with the above, and justify the name \emph{Martingale Sinkhorn Algorithm}, note that we can rewrite 
the relation in \cref{fig:MS_algorithm} and \cref{alg:MSinkhorn} as 
\begin{eqnarray}
& \qquad \text{\emph{Martingale Schr\"odinger system}} \qquad & \qquad \text{\emph{Martingale Sinkhorn Algorithm}}\nonumber\\
&\mu_0 = (F*\gamma)_\# \alpha &\text{ Update } \alpha \qquad \mu_0 = (F_{i-1}*\gamma)_\# \alpha_i,\label{eq:MSsystem}\\
&\mu_1 = F_\# (\alpha* \gamma)  &\text{ Update } F \qquad \mu_1 = (F_i)_\# (\alpha_i* \gamma),\nonumber
\end{eqnarray}
where we wrote $F=\nabla v^\FL$ the gradient of the FL transform of the convex potential $v$.

\subsection{Related Literature} \label{subsec:literature}
In one dimension, \cite{conze2021bass} proposed a fixed-point algorithm to solve the martingale Benamou-Brenier problem \eqref{eq:dmbb}. 
It is insightful to comment on the link between \cref{alg:MSinkhorn} and their approach. The key simplification in one-dimension is that the Brenier-McCann potential is given explicitly, namely $\nu = (G_\nu^{-1}\circ G_\mu)_\# \mu$ with $G_\mu$ the cumulative distribution function of $\mu$.  This renders both relations in the Martingale Schr\"odinger system \eqref{eq:MSsystem} explicit, namely 
\begin{equation*}
\gamma * F = G_{\mu_0}^{-1} \circ G_{\alpha}\quad \textrm{and}\quad F= G_{\mu_1}^{-1}\circ G_{\gamma * \alpha}.
\end{equation*}
Note also that $G_{\gamma * \alpha} = \gamma*G_{\alpha}$ so that we obtain
$$ G_{\alpha} = G_{\mu_0}\circ \left(\gamma * F\right) = G_{\mu_0}\circ \left(\gamma * \left( G^{-1}_{\mu_1}\circ \left(\gamma*G_{\alpha}\right)   \right) \right)$$
which is the fixed point relation for $G_{\alpha}$ in \cite{conze2021bass}. \cite{acciaio2025calibration} showed that the resulting fixed-point algorithm converges, but only under strong regularity assumptions that $
mu_1$ is compactly supported with a density bounded away from zero.  Numerically, it is very quick and can be used to calibrate a one-dimensional financial market model to option prices at multiple maturities, as shown in \cite{conze2021bass}. In addition, \cite{gMBB} recently considered \eqref{eq:mbb} but with geometric Brownian motion as the reference process and obtained an explicit algorithm to solve it, see also \cite{beiglbock2025change}. In one dimension, our work complements these results, establishes convergence under minimal assumptions and offers new insight, in particular through the analogy to the classical Sinkhorn algorithm, as discussed above.

In higher dimensions, little was known about how to compute the Bass potential and the associated Bass measure solving~\eqref{eq:dmbb} and our work offers, to the best of our knowledge, the first general algorithm to solve this challenge. \cite{backhoff2025gradient_flow} introduced the lifted functional
\begin{equation*}
    V(Z) := \sup_{\substack{
        Y \in L^2(\mathbb{R}^d;\sigma(X,\Gamma)),\\
        Y \sim \mu_1
    }}
    \mathbb{E}\!\left[(Z+\Gamma)Y - ZX\right], \qquad Z \in L^2(\mathbb{R}^d;\sigma(X)),
\end{equation*}
where $\Gamma \sim \gamma$ and $X \sim \mu_0$. They demonstrate that, for compactly supported marginals admitting a Bass measure with finite second moment, this lifted formulation is equivalent to~\eqref{eq:dmbb} and that its $L^2$-gradient flow is contractive. More directly relevant to our approach is the work \cite{backhoff2023bass} where the authors introduced what they called the \emph{Bass functional}  
\[
    \mathcal{V}(\alpha)
    := \MCov(\alpha * \gamma, \mu_1) - \MCov(\alpha, \mu_0),
    \qquad \alpha \in \mathcal{P}_2(\mathbb{R}^d).
\]
They show that
\[
    \mathbf{P}_{\mu_0,\mu_1}
    = \inf_{\alpha \in \mathcal{P}_2(\mathbb{R}^d)} \mathcal{V}(\alpha),
\]
and that the infimum is attained by $\hat \alpha \in \mathcal{P}_2(\R^d)$ if and only if $\hat \alpha$ is the Bass measure associated with $(\mu_0,\mu_1)$. As discussed in \cref{subsec:main_results}, the Martingale Sinkhorn Algorithm strictly decreases the dual objective $\Ecal$ at each iteration. In particular, when the marginals are compactly supported, \cref{lem:strict_descent_compact} shows that this is achieved by simultaneously increasing $\MCov(\mu_0,\cdot)$ and decreasing $\MCov(\cdot,\mu_1)$. By the same reasoning, one obtains
\[
    \mathcal{V}(\alpha_{i+1}) < \mathcal{V}(\alpha_{i})
\]
for the iterates $(\alpha_i)_{i\in\N}$ of the algorithm, as long as $\alpha_i$ is not the Bass measure.

\section{Convergence of the Martingale Sinkhorn Algorithm} \label{sec:convergence}

We now turn to the convergence analysis of \cref{alg:MSinkhorn}. Our approach follows the dual formulation of the Martingale Benamou--Brenier problem considered in \cite{backhoffveraguas2025existence}, and is based on the study of the dual objective functional
\[
    \Ecal(v) = \int v\,\mathrm{d}\mu_1 - \int v^C\,\mathrm{d}\mu_0.
\]
Throughout this section we set $I_{\mu_1} := \operatorname{ri}\bigl(\co(\supp(\mu_1))\bigr)$. Since $I_{\mu_1}$ is relatively open in its affine hull, we use the standard interior notation for balls and neighborhoods, interpreted in the trace topology. Moreover, under our irreducibility assumption we have that $\mu_0$ does not charge the boundary of $\co(\supp(\mu_1))$. Indeed, if it did, any connecting martingale would have to send mass from the boundary into the interior, which is impossible because a boundary point admits a supporting hyperplane that strictly separates it from the interior. Hence $\mu_0(I_{\mu_1}) = 1$.

We begin by collecting several elementary properties of $\Ecal$. To this end, for any finite measure $\eta$,
we define
\[
    \bar\eta := \frac{1}{\eta(\R^d)} \int x\,\eta(\rmd x)
\]
the barycenter of $\eta$, whenever it exists. In particular, our pair $(\mu_0,\mu_1)$ satisfies $\bar\mu_0 = \bar\mu_1$.

\begin{lemme} \label{lem:obj_props}
    Let $v\in L^1(\mu_1)$ be l.s.c.~and convex, and assume that $v^{\FL}*\gamma$ is proper. Then $v^C\in L^1(\mu_0)$ and $\Ecal(v)$ is well-defined with $\Ecal(v)\ge0$.
\end{lemme}

\begin{proof}
    Since $v^{\FL} * \gamma$ is proper, let $y_0 \in \R^d$ with $(v^{\FL} * \gamma)(y_0) < \infty$. By the Fenchel--Young inequality, for all $x \in \R^d$, 
    \[
        v^C(x) \geq y_0 \, x- (v^{\FL} * \gamma)(y_0).
    \]
    Jensen's inequality gives $v^{\FL} * \gamma \geq v^{\FL}$, hence $v^C\leq v$. Since $v \in L^1(\mu_1)$, we get $(v^C)^+ \in L^1(\mu_1)$. The affine lower bound above and finite first moments of $\mu_0$, $\mu_1$ yield $(v^C)^- \in L^1(\mu_1) \cap L^1(\mu_0)$. As $\mu_0 \leq_{\rm cx} \mu_1$, it follows that $(v^C)^+ \in L^1(\mu_0)$. Thus ${v^C \in L^1(\mu_0)}$, so $\Ecal(v)$ is well-defined and since $v^C\leq v$, we get $\Ecal(v) \geq 0$.
\end{proof}

\medskip

Since Bass potentials need not be integrable with respect to $\mu_1$, we work with an extended version of the objective: For $\pi\in\Cpl_M(\mu_0,\mu_1)$ define,
for any l.s.c.~convex $v$ such that $v^\star*\gamma$ is proper,
\[
    \Ecal_\pi(v) := \int\!\left(\int v\,\rmd\pi_x -v^C(x)\right)\,\mu_0(\rmd x).
\]
By Jensen's inequality and the definition of the $C$-transform, for $\mu_0$-a.e.~$x$,
\[
    \int v\,\rmd\pi_x-v^C(x)\ge v(x)-v^C(x)\ge 0,
\]
hence $\Ecal_\pi(v)\in[0,\infty]$. If $v\in L^1(\mu_1)$ and $v^C\in L^1(\mu_0)$, then Fubini and the marginal constraint yield $\Ecal_\pi(v)=\Ecal(v)$.

\begin{lemme} \label{lem:ext_obj_props}
    Let $\mu_0,\mu_1\in\Pcal_p(\Rd)$, $\pi\in\Cpl_M(\mu_0,\mu_1)$, and let $v$ be l.s.c.\ convex. Assume that $\Ecal_\pi(v)<\infty$ and that $(\nabla v^\star)_\#(\alpha*\gamma)=\mu_1$ for some $\alpha\in\Pcal(\Rd)$. Then there exists an affine map $\ell$ such that $(v+\ell)^C\ge 0$ and $(v+\ell)^C(\bar\mu_0)=0$.
    Moreover, $\Ecal_\pi(v)$ is invariant under affine shifts. 
\end{lemme}

\begin{proof}
    Since $(\nabla v^\star)_\#(\alpha*\gamma)=\mu_1$, we have
    \[
        \int |\nabla v^\star|^p\,\rmd(\alpha*\gamma) =\int |x|^p\,\rmd\mu_1(x)<\infty.
    \]
    In particular, there exists $y\in\Rd$ such that $\int |\nabla v^\star(y+z)|^p\,\rmd\gamma(z)<\infty$, and by \cref{lem:exp.bound} we obtain $v^\star\in L^1(\gamma_y)$ for all $y\in\Rd$. Hence $v^\star*\gamma$ is proper and $\operatorname{dom}(v^\star*\gamma) = \Rd$, and therefore $v^C=(v^\star*\gamma)^\star$ is proper as well.
    Since $\Ecal_\pi(v)<\infty$, the quantity $\int v\,\rmd\pi_x - v^C(x)$ is finite for $\mu_0$-a.e.~$x$. As $\int v\,\rmd\pi_x \ge v^C(x)$ $\mu_0$-a.e., we deduce that $\int v\,\rmd\pi_x<\infty$ and $v^C(x)<\infty$ for $\mu_0$-a.e.~$x$. Moreover, $\pi_x(v=\infty)=0$ for $\mu_0$-a.e.~$x$, hence $\mu_1(v=\infty)=0$. By convexity, it follows that $v<\infty$ on $I_{\mu_1}$, i.e., $I_{\mu_1}\subset \operatorname{dom}(v)$. Since $v^C\le v$, we also have $I_{\mu_1}\subset \operatorname{dom}(v^C)$.
    
    If $\operatorname{dom}(v)$ is not a singleton then $\partial v^\star(\Rd)$ is not a singleton, and $v^\star*\gamma$ is strictly convex on $\Rd$ (see, e.g., \cite[Proposition 5.2]{backhoffveraguas2025existence} and the proof thereof). Therefore, by \cite[Theorem 26.3]{R}, $v^C$ is differentiable on $\operatorname{int}(\operatorname{dom}(v^C)) \neq \emptyset$. In particular, ${I_{\mu_1} \subset \operatorname{int}(\operatorname{dom}(v^C))=\partial(v^\star*\gamma)(\Rd)}$. As $\bar\mu_0\in I_{\mu_1} $, there exists $y\in\Rd$ such that $\bar\mu_0\in\partial(v^\star*\gamma)(y)$. Define the affine map $\ell(x):=(v^\star*\gamma)(y)- y\,x$. Then the Fenchel--Young inequality yields $v^C+\ell\ge 0$ and $(v^C+\ell)(\bar\mu_0)=0$. If $\operatorname{dom}(v) = \{\bar\mu_0\}$, then $v^C(\bar\mu_0) \in \R$ and $+\infty$ elsewhere. With $\ell \equiv -v^C(\bar\mu_0)$, we have $v^C+\ell \geq 0$ and $(v^C+\ell)(\bar\mu_0) = 0$.
    
    Finally, if $\ell$ is affine, then $(v+\ell)^C=v^C+\ell$, and therefore
    \[
        \Ecal_\pi(v+\ell) =\int\!\left(\int v\,\rmd\pi_x + \ell(x)- (v+\ell)^C(x)\right)\,\mu_0(\rmd x) =\Ecal_\pi(v). \qedhere
    \]
\end{proof}

\medskip

In view of \cref{lem:ext_obj_props}, we normalise the iterates of \cref{alg:MSinkhorn} by affine shifts. More precisely, for each $i\in\N$ we choose an affine map $\ell_i$ such that
\[
    (v_i+\ell_i)^C \ge 0, \quad (v_i+\ell_i)^C(\bar\mu_0)=0,
\]
and set $\hat v_i:=v_i+\ell_i$. Writing $(\nabla v_i)_\#\mu_1=\alpha_i*\gamma$, we then have
\[
    (\nabla \hat v_i)_\#\mu_1=\hat\alpha_i*\gamma, \quad \hat\alpha_i := (\operatorname{id}+\nabla \ell_i)_\#\alpha_i.
\]
Accordingly, the convergence analysis is carried out for the normalised sequence $(\hat v_i)_{i\in\N}$.
\medskip

We state now the standing assumptions on $\mu_0,\mu_1$, and the initial potential $v_0$ under which we prove convergence of \cref{alg:MSinkhorn}:
\begin{enumerate}
    \asptitem{A1}{A1} \label{aspt:irreducible}
    $\mu_0, \mu_1 \in \mathcal{P}_{p}(\mathbb{R}^d)$, for some $p > 1$, such that $\mu_0$ and $\mu_1$ are in convex order and irreducible;
    
    \asptitem{A2}{A2} \label{aspt:integrable}
    $v_0 \in L^1(\mu_1)$ is l.s.c.~convex;
    
    \asptitem{A3}{A3} \label{aspt:proper}
    $v_0^{\FL} * \gamma$ is proper;
\end{enumerate}

In addition, throughout this section, and especially in the general formulation of our convergence result, we fix once and for all a martingale coupling ${\pi'\in\Cpl_M(\mu_0,\mu_1)}$ satisfying the geometric properties stated below.

\begin{prop}[Existence of a super-spreader]\label{prop:existence_super-spreader}
    If $\mu_0$ and $\mu_1$ are in convex order and irreducible, there exists $\pi'\in\Cpl_M(\mu_0,\mu_1)$ such that, for $\mu_0$-almost every $x$, $\pi'_x(U) > 0$ for all open sets $U \subset\Rd$ with $\mu_1(U) >0$. In addition, any Bass martingale coupling~\eqref{eq:Basscoupling} satisfies this property.
    
    In particular, there exists $\delta\in(0,1)$, $x_1,\dots,x_m\in\supp(\mu_1)$, such that
    \begin{align*}
        & \overline{B}_\delta(\bar\mu_0)\subset \interior(\co\{y_1,\dots,y_m\})
        &&
        \forall\, y_i\in B_\delta(x_i),\ \forall \,i=1,\dots,m, \\
        & \pi'_x\bigl(B_\delta(x_i)\bigr)>0 
        &&
        \text{for $\mu_0$-a.e.\ }x,\ \forall\, i=1,\dots,m, \\
        & B_\delta(x_i)\cap B_\delta(x_j)=\emptyset
        &&
        \forall\, i\neq j.
    \end{align*}
\end{prop}

\begin{proof}
    We defer the proof of this proposition to \cref{sec:appendix:auxiliary_results}.
\end{proof}

In our tightness argument, this construction guarantees that blow-ups of the convex potentials obtained from \cref{alg:MSinkhorn} cannot occur solely on $\pi'_x$-negligible sets. Indeed, each $\pi'_x$ assigns positive mass to disjoint regions whose convex hull covers a neighbourhood of the reference point $\bar\mu_0$.

\medskip 
To begin with, we establish strict descent of the objective under an additional compactness requirement. This compact case will not be used later (in the general analysis we rely on \cref{lem:strict_descent}), but it illustrates in a direct way how the objective decreases along the algorithm:
\begin{enumerate}
    \asptitem{AC}{AC} \label{aspt:compact_support} 
    There exists a compact $K_0 \subset \R^d$ with $\supp (\mu_0) \subseteq K_0 \subset I_{\mu_1}$;
\end{enumerate}

As a preliminary step, we show that under \ref{aspt:irreducible} and \ref{aspt:compact_support} the Martingale Sinkhorn algorithm preserves Assumptions~\ref{aspt:integrable} and~\ref{aspt:proper} throughout its iterations. Under these conditions, \cref{lem:strict_descent_compact} shows that $\Ecal_{\pi'}$ decreases strictly until the Bass measure is attained.

\begin{lemme} \label{lem:preservation_struc_props}
    Under Assumptions \ref{aspt:irreducible}, \ref{aspt:compact_support}, \ref{aspt:integrable} and \ref{aspt:proper}, \cref{alg:MSinkhorn} yields compactly supported measures $(\alpha_i)_{i \in \N}$ and potentials $(v_i)_{i \in \N}$ which satisfy \ref{aspt:integrable} and \ref{aspt:proper}.
\end{lemme}

\begin{proof}
    By \ref{aspt:proper}, we have $v_0^C=(v_0^{\FL} * \gamma)^{\FL} \leq v_0$, hence $\operatorname{dom}\bigl(v_0^C\bigr) \supseteq \operatorname{dom}(v_0) \supseteq I_{\mu_1}$ by \ref{aspt:integrable}. Under \ref{aspt:compact_support}, there exists a compact $K_0 \subset \R^d$ with $\mu_0(K_0) = 1$ and $K_0 \subset I_{\mu_1}$. Since $(v_0^{\FL} * \gamma)^{\FL}$ is Lipschitz on $K_0$ (see \cite[Theorem 10.4]{R}), $\alpha_{1} := \nabla ( v_{0}^{\FL} * \gamma)^{\FL}_\# \mu_0$ is well-defined and compactly supported. As $v_1$ attains $\MCov(\alpha_{1} * \gamma, \mu_1)$, it follows that $v_1 \in L^1(\mu_1)$ and $v_1^{\FL} * \gamma$ is $\alpha_1$-integrable, hence proper, which remains valid upon affine shifts. Repeating this argument for each iterate shows the claim holds for all $i$ by induction.
\end{proof}

\begin{lemme}[Strict Descent under Compact Support] \label{lem:strict_descent_compact}
    Under Assumptions \ref{aspt:irreducible}, \ref{aspt:compact_support}, \ref{aspt:integrable} and \ref{aspt:proper}, \cref{alg:MSinkhorn} strictly decreases $\Ecal$ in every step $i \in \N$ as long as $\alpha_{i}$ is not a Bass measure, i.e., $\alpha_{i} \neq \alpha_{i+1}$ if and only if $\Ecal(v_{i}) < \Ecal(v_{i-1})$.
\end{lemme}
\begin{proof}  
    By \cref{lem:preservation_struc_props}, we can write
    \[
         \Ecal(v_{i-1})=\int v_{i-1}\,\mathrm{d}\mu_1 - \int v_{i-1}^C \,\mathrm{d}\mu_0 = \underbrace{\left(\int v_{i-1}^{\FL}*\gamma \,\mathrm{d}\alpha_{i} + \int v_{i-1}\,\mathrm{d}\mu_1\right)}_{\displaystyle \rm (I)} - \underbrace{\left(\int v_{i-1}^C \,\mathrm{d}\mu_0 + \int v_{i-1}^{\FL}*\gamma\,\mathrm{d}\alpha_{i}\right)}_{\displaystyle \rm (II)}.
    \]
    Since $v_i$ attains $\MCov(\alpha_i*\gamma,\mu_1)$ by construction, we have
    \[
        {\rm (I)} \geq \MCov(\alpha_i * \gamma, \mu_1) = \int v_i^{\FL}*\gamma \,\mathrm{d}\alpha_i + \int v_i\,\mathrm{d}\mu_1.
    \]
    Likewise, $v_{i-1}^C$ achieves $\MCov(\mu_0, \alpha_i)$, hence
    \[
        {\rm (II)} = \MCov(\mu_0, \alpha_i) \leq \int v_{i}^C \, \mathrm{d}\mu_0 + \int v_i^{\FL} * \gamma \,\mathrm{d}\alpha_i.
    \]
    Therefore,
    \begin{align*}
         {\rm (I) -(II)} &\geq \MCov(\alpha_{i} * \gamma, \mu_1) - \left(\int v_{i}^C \,\mathrm{d}\mu_0 + \int v_i^{\FL} * \gamma\,\mathrm{d}\alpha_i\right) = \int v_{i}\,\mathrm{d}\mu_1 - \int v_{i}^C \,\mathrm{d}\mu_0 = \Ecal(v_{i}).
    \end{align*}
    The inequality is either strict, or else we have ${\rm (I)} = \MCov(\alpha_i * \gamma, \mu_1)$, hence $(\nabla v_{i})_\# \mu_1 = \alpha_{i} * \gamma = (\nabla v_{i-1})_\# \mu_1$. Since $\alpha_{i} := (\nabla v_{i-1}^C)_\# \mu_0$ by construction, $\alpha_i$ already is a Bass measure.
\end{proof}

To extend \cref{lem:strict_descent_compact} to the non-compact case, we recall the notion of epi-convergence and a few basic consequences for convex potentials: Let $E \subset \Rd$ be open and non-empty, and let $(f_n)_{n\in\N}$, $f$ be functions $E \to \R \cup\{+\infty\}$. We say that $f_n \to f$ in epi-convergence if, for every $x \in E$,
\begin{align*}
    f(x) &\leq \liminf_{n\to\infty} f_n(x_n) \text{ for every sequence } x_n \to x,\\
    f(x) &\geq \limsup_{n\to\infty} f_n(x_n) \text{ for some sequence } x_n \to x.
\end{align*}
We collect some standard properties of epi-convergence for convex functions.

\begin{lemme}[Properties of epi-convergence]\label{lem:properties_epi_convergence}
Let $(f_n)_{n\in\N}$ be proper l.s.c.~convex functions $E \to \R \cup\{+\infty\}$.
\begin{enumerate}
    \item If $f_n \rightarrow f$ in epi-convergence, then $f$ is l.s.c.~and convex (possibly improper).
    \item If $(f_n)_{n \in \N}$ does not epi-converge to the identically $+\infty$ function, then there exist a subsequence $(f_{n_k})_{k \in \N}$ and a function $f \not\equiv +\infty$ such that $f_{n_k} \to f$ in epi-convergence.
    \item Let $f$ be l.s.c.~convex with $\operatorname{int}(\operatorname{dom} f) \neq \emptyset$. Then the following are equivalent:
    \begin{enumerate}
        \item $f_n \to f$ in epi-convergence;
        \item $f_n \to f$ uniformly on every compact set $K \subset E$ with $K \cap \partial(\operatorname{dom} f) = \emptyset$, where $\partial(\operatorname{dom} f)$ is the boundary of $\operatorname{dom} f$;
        \item there exists a dense set $D \subset E$ such that $f_n(x) \to f(x)$ for all $x \in D$.
    \end{enumerate}
    \item Let $f$ be proper l.s.c.~convex. Then $f_n \to f$ in epi-convergence if and only if $f_n^\star \to f^\star$ in epi-convergence.
    \item If $f_n \to f$ in epi-convergence, then for every $(x,z)$ with $z \in \partial f(x)$ there exist sequences $(x_n,z_n)$ such that $z_n \in \partial f_n(x_n)$ and $(x_n,z_n) \to (x,z)$.
\end{enumerate}
\end{lemme}

\begin{proof}
All statements are standard; see \cite{rockafellar2009variational}. 
Specifically, (1) follows from Proposition~7.4 and the first part of Theorem~7.17, 
(3) is the second part of Theorem~7.17, (2) follows from Theorem~7.6, 
(4) is Theorem~11.34, and (5) follows from Theorem~12.35.
\end{proof}

The previous lemma has two important consequences:
\begin{enumerate}
    \item If $(f_n)_{n\in\N}$ is a sequence of l.s.c.~convex functions that is locally bounded, then by Lemma~\ref{lem:properties_epi_convergence} (1) and (2) there exists a subsequence that epi-converges to an l.s.c.~convex function $f$. Local boundedness also prevents $f$ from taking the values $\{\pm\infty\}$, so $f$ is proper.
    \item Let $f_n \to f$ in epi-convergence with $\operatorname{int}(\operatorname{dom} f) \neq \emptyset$, and let $C \subset \operatorname{int}(\operatorname{dom} f)$ be a bounded open convex set. Then $f_n \to f$ uniformly on $\overline C$ by Lemma~\ref{lem:properties_epi_convergence}(3), hence pointwise on $C$. Proper l.s.c.~convex functions are locally Lipschitz on the interior of their domain, so $f$ and each $f_n$ are differentiable almost everywhere on $C$ (by Rademacher's theorem). Combining this with \cite[Theorem 24.5]{R}, applied to the restrictions of $f_n$ and $f$ to $C$, we obtain $\nabla f_n(x) \to \nabla f(x)$ for a.e.~$x \in C$.
\end{enumerate}

\medskip

We are now in a position to extend \cref{lem:strict_descent_compact} to the non-compact case. In a nutshell, this result shows that as long as a potential $v$ is not yet a Bass potential, one iteration of \cref{alg:MSinkhorn} strictly decreases the objective. \Cref{cor:the_universal_gap} then establishes that this improvement is stable under epi-convergence.

\begin{lemme}[Strict Descent] \label{lem:strict_descent}
    Let Assumption \cref{aspt:irreducible} hold. Let $\alpha\in \Pcal(\R^d)$ and $v,w$ be l.s.c.~convex functions with $(\nabla v^C)_\# \mu_0 = \alpha$ and $(\nabla w^{\FL})_\# (\alpha * \gamma) = \mu_1$. Moreover, assume that $v^\FL*\gamma$ is proper.
    If there exist $\epsilon, \delta > 0$ such that
    \begin{equation}
        \label{eq:probability_gap}
        \alpha * \gamma(\{ v \circ \nabla w^{\FL} + v^{\FL} > w \circ \nabla w^{\FL} + w^{\FL} + \epsilon\}) = \delta > 0,
    \end{equation}
    then
    \[
        \Ecal_{\pi'}(w) \leq \Ecal_{\pi'}(v) - \epsilon \delta .
    \]
    Furthermore, if $\Ecal_{\pi'}(v) < \infty$ and $v-w$ is not constant, there exist $\epsilon,\delta > 0$ satisfying \eqref{eq:probability_gap}.
\end{lemme}

\begin{proof}
    Assume $\Ecal_{\pi'}(v) < \infty$ as otherwise, there is nothing to prove. Let $(K_n)_{n \in \N}$ be compacts with $K_n \uparrow I_{\mu_1}$, $\mu_0(K_n) > 0$, and set $\mu_0^{(n)} := \mu_0(\cdot \cap{K_n})$. By \cref{lem:ext_obj_props} and the proof thereof, $I_{\mu_1} \subset \operatorname{dom}(v^C)$ and hence $v^C \in L^1\bigl(\mu_0^{(n)}\bigr)$. With $\mu_1^{(n)} := \int \pi_x'\,\mu_0^{(n)}(\rmd x)$ we have, for all $n \in \N$,
    \[
        \int \!\left(\int v\,\rmd\pi_x' -v^C(x)\right) \mu_0^{(n)}(\rmd x) \leq \Ecal_{\pi'}(v) < \infty,
    \]
    and therefore $v \in L^1\bigl(\mu_1^{(n)}\bigr)$. It follows that $\mu_0^{(n)}$, $\mu_1^{(n)}$ and $v$ satisfy \cref{aspt:irreducible}, \cref{aspt:compact_support}, \cref{aspt:integrable} and \cref{aspt:proper} for all $n \in \N$. 
    Set $\alpha^{(n)} := (\nabla v^C)_\# \mu_0^{(n)}$ and let $w_n$ be l.s.c.~convex such that $(\nabla w_n^{\FL})_\# (\alpha^{(n)} * \gamma) = \mu_1^{(n)}$. By construction, for all $n \in \N$,
    \[
        \int |\nabla w_n^\FL|^p \, \rmd(\alpha^{(1)}*\gamma) = \int \left|\nabla w_n^\FL(\nabla v^C(x)+z)\right|^p \rmd\mu_0^{(1)}\!\otimes\gamma(x,z) \leq \int |\nabla w_n^\FL|^p \, \rmd\alpha^{(n)}*\gamma = \int |x|^p\, \mu_1^{(n)}(\rmd x),
    \]
    and therefore, after extraction of a subsequence, there exists $y_0 \in \Rd$ with
    \begin{equation} \label{eq:lem:strict_descent:bnded_grads}
        \sup_{n \in \N}\int |\nabla w_n^\FL(y_0+z)|^p \, \rmd\gamma(z) < \infty.
    \end{equation}
    Since $(w_n)_{n \in \N}$ satisfy \cref{aspt:integrable} and \cref{aspt:proper} by \cref{lem:preservation_struc_props}, we may normalise $(w_n^\FL)_{n \in \N}$ so that, for all $n \in \N$,
    \begin{equation} \label{eq:lem:strict_descent:normalisation}
        c := \int_{B_1(y_0)} w^\FL(x)\,\frac{\rmd x}{|B_1(y_0)|}=\int_{B_1(y_0)} w_n^\FL(x)\,\frac{\rmd x}{|B_1(y_0)|}.
    \end{equation}
    In particular, 
    \[
        \sup_{n \in \N}\int_{B_1(y_0)}|\nabla w_n^\FL(x)|^p \, \rmd x < \infty,
    \]
    and by the Poincar\'e-Wirtinger inequality there is $C_p > 0$ such that
    \[
        \sup_{n \in \N}\int_{B_1(y_0)} |w_n^\FL(x) - c|^p\,\rmd x \leq C_p \, \sup_{n \in \N}\int_{B_1(y_0)} |\nabla w_n^\FL(x)|^p\,\rmd x < \infty.
    \]
    Let $W^{1,p}(B_1(y_0))$ be the Sobolev space of all $f \in L^p(B_1(y_0))$ such that
    \[
      \lVert f\rVert_{W^{1,p}(B_1(y_0))} := \lVert f\rVert_{L^p(B_1(y_0))} + \lVert \nabla f\rVert_{L^p(B_1(y_0))} < \infty.
    \]
    By the Rellich–Kondrachov theorem $W^{1,p}(B_1(y_0)) \subset L^p(B_1(y_0))$ is compact, and since 
    \[
        \sup_{n\in\mathbb{N}}\|w_n^\FL\|_{W^{1,p}(B_1(y_0))} < \infty, 
    \]
    there exists a subsequence (still denoted $(w_n^\FL)_{n \in \N}$) such that $(w_n^\FL)_{n \in \N}$ converges in $L^p(B_1(y_0))$. Passing to a further subsequence, we may assume that $(w_n^\FL)_{n \in \N}$ converges almost everywhere on $B_1(y_0)$. By \cref{lem:properties_epi_convergence} and subsequent remarks, its limit is a finite convex function on $B_1(y_0)$ and the convergence is locally uniform on $B_1(y_0)$. In particular, there exists $\delta \in (0,1)$ such that ${\sup_{n\in\N} |w_n^\star| < \infty}$ on $B_\delta(y_0)$. Together with \eqref{eq:lem:strict_descent:bnded_grads}, \cref{lem:exp.bound} implies ${\sup_{n \in \N} |w_n^\star| < \infty}$ on $\Rd$. By \cref{lem:properties_epi_convergence} and the subsequent remarks, passing to a further subsequence yields an l.s.c.~convex $\tilde w^\FL$ such that $w_n^\FL \to \tilde w^\FL$ in epi-convergence and $\nabla w_n^\FL \to \nabla\tilde w^\FL$ $\alpha*\gamma$-almost surely. Since $\alpha^{(n)} \to \alpha$ and $\mu_1^{(n)} \to \mu_1$ in total variation, it follows that $\tilde w^\FL$ is a Brenier--McCann potential for $(\alpha * \gamma,\mu_1)$ (see, e.g., \cref{lem:limit_potential}), and so $\tilde w^{\FL}$ and $w^{\FL}$ differ at most by an additive constant. Indeed, $\tilde w^\FL = w^\FL$ by \eqref{eq:lem:strict_descent:normalisation}.    
    
    Using Fenchel--Young's inequality, 
    \begin{align*}
        \Ecal_{\pi'}(v) &= \lim_{n \to \infty} \int \!\left(\int v \,\mathrm{d}\pi'_x - v^C(x)\right) \mu_0^{(n)}(\rmd x)
        \\
        &=\lim_{n \to \infty} \left[\int v \circ \nabla w_n^{\FL} + v^{\FL} \, \mathrm{d}(\alpha^{(n)} * \gamma) - \int v^C \circ \nabla (v^{\FL} * \gamma) + v^{\FL} * \gamma  \, \mathrm{d}\alpha^{(n)}\right]
        \\
        &\geq
        \lim_{n \to \infty} \left[\int w_n \circ \nabla w_n^{\FL} + \epsilon \mathbbm 1_{\{ v \circ \nabla w_n^{\FL} + v^{\FL} > w_n \circ \nabla w_n^{\FL} + w_n^{\FL} + \epsilon \} } \, \mathrm{d}(\alpha^{(n)}*\gamma) - \int w^C_n \circ \nabla(v^{\FL}*\gamma) \, \mathrm{d}\alpha^{(n)}\right]
        \\
        &=
        \lim_{n \to \infty} \left[ \epsilon \alpha^{(n)} * \gamma(\{ v \circ \nabla w_n^{\FL} + v^{\FL} > w_n \circ \nabla w_n^{\FL} + w_n^{\FL}+ \epsilon \}) + \int\!\left(\int w_n \,\mathrm{d}\pi'_x - w_n^C(x) \right)\mathrm{d}\mu_0^{(n)}(x)\right]
        \\
        &\geq \epsilon \liminf_{n \to \infty} \left[\alpha^{(n)} * \gamma(\{ v \circ \nabla w_n^{\FL} + v^{\FL} > w_n \circ \nabla w_n^{\FL} + w_n^{\FL} + \epsilon\})\right] + \Ecal_{\pi'}(w).
    \end{align*}
    The last inequality follows from Fatou’s lemma and the inequality $w_n^C(x) \leq \int w_n\,\mathrm{d}\pi'_x$. 
    Since $w^{\FL}_n \to w^{\FL}$ pointwise and $\nabla w_n^{\FL} \to \nabla w^{\FL}$ $\alpha*\gamma$-almost surely, we have for almost every $z$,
    \[
        \lim_{n \to \infty} w_n \circ \nabla w_n^{\FL}(z) + w_n^{\FL}(z) = \lim_{n \to \infty} \nabla w_n^{\FL}(z) \cdot z = \nabla w^{\FL}(z) \cdot z = w \circ \nabla w^{\FL} + w^{\FL},
    \]
    Because $v$ is l.s.c.~convex, we also have for almost every $z$,
    \[
        \liminf_{n \to \infty} v \circ \nabla w_n^{\FL}(z) + v^\FL (z) \ge v \circ \nabla w^{\FL}(z) + v^\FL(z).
    \]
    Therefore,
    \begin{multline*}
        \liminf_{n \to \infty}
        \alpha^{(n)}*\gamma(\{ v \circ \nabla w_n^{\FL} + v^{\FL} > w_n \circ \nabla w_n^{\FL} + w_n^{\FL} + \epsilon \}) 
        \\
        =\liminf_{n \to \infty}
        \alpha * \gamma(\{ v \circ \nabla w_n^{\FL} + v^{\FL} > w_n \circ \nabla w_n^{\FL} + w_n^{\FL} + \epsilon\})
        \\
        \ge~\alpha * \gamma(\{ v \circ \nabla w^{\FL} + v^{\FL} > w \circ \nabla w^{\FL} + w^{\FL} + \epsilon\}) \ge \delta,
    \end{multline*}
    and hence $\Ecal_{\pi'}(w) \leq \Ecal_{\pi'}(v) - \epsilon\delta $.
    
    For the final claim, \cref{lem:disjoint_subdiffs} yields an open $B \subseteq \R^d$ with ${\partial v^{\FL}(z) \cap \partial w^{\FL}(z) = \emptyset}$ for all $z \in B$.
    Therefore for almost every $z \in B$,
    \[
        v \circ \nabla w^{\FL}(z) + v^{\FL}(z) > z \, \nabla w^{\FL}(z) = 
        w \circ \nabla w^{\FL}(z) + w^{\FL}(z).
    \]
    Since $\gamma * \alpha(B) > 0$, there exists $\epsilon > 0$ such that
    \[
        \alpha * \gamma( \{ v \circ \nabla w^{\FL} + v^{\FL} >
        w \circ \nabla w^{\FL} + w^{\FL} + \epsilon \} ) \ge \alpha * \gamma(B) / 2 > 0. \qedhere
    \]
\end{proof}

\begin{cor}\label{cor:the_universal_gap}
    Let $v, w, \alpha$ be as in \cref{lem:strict_descent}. Let $(v_i)_{i\in\N}$ and $(w_i)_{i\in\N}$ be sequences of l.s.c.~convex functions with $(v_i)_{i\in\N} \to v$ and $(w_i)_{i\in\N} \to w$ in epi-convergence on $I_{\mu_1}$ $(\alpha_i)_{i\in\N}$ be a sequence of probability measures with $(\alpha_i)_{i \in \N} \to \alpha$ in total variation such that $(\nabla v_i^C)_\# \mu_0 = \alpha_i$ and $(\nabla w_i^{\FL})_\# (\alpha_i * \gamma) = \mu_1$ for all $i \in \N$. Then there exist $\epsilon > 0$ and $i_0 \in \N$ such that, for all $i \geq i_0$,
    \[
        \Ecal_{\pi'}(w_i) \leq \Ecal_{\pi'}(v_i) - \epsilon.
    \]
\end{cor}

\begin{proof}
    \cref{lem:strict_descent} ensures existence of $\epsilon, \delta > 0$ such that
    \[
        \alpha * \gamma\bigl(\{v \circ \nabla w^{\FL} + v^{\FL} > w \circ \nabla w^{\FL} + w^{\FL} + \epsilon\}\bigr) = 4\delta > 0,
    \]
    and $v_i \to v$ and $w_i \to w$ in epi-convergence implies $v_i^{\FL} \to v^{\FL}$ and $w_i^{\FL} \to w^{\FL}$ pointwise on $\R^d$. By \cref{lem:properties_epi_convergence} and subsequent remarks, we have $\nabla w_i^{\FL} \to \nabla w^{\FL}$ and hence, by epi-convergence of $(v_i)_{i\in \N}$,
    \[
        \liminf_{i\to\infty} v_i \circ \nabla w^\star_i \geq v \circ \nabla w^\star
    \]
    $\alpha * \gamma$-almost surely. Using also the Fenchel--Legendre duality, which gives $w_i\circ\nabla w_i^{\FL} + w_i^{\FL} = \operatorname{id} \nabla w_i^{\FL}$ for all $i \in \N$, we obtain
    \[
        \liminf_{i \to \infty} \alpha * \gamma\bigl(\{v_i \circ \nabla w_i^{\FL} + v_i^{\FL} > w_i \circ \nabla w_i^{\FL} + w_i^{\FL} + \epsilon\}\bigr) \geq \alpha * \gamma\bigl(\{v \circ \nabla w^{\FL} + v^{\FL} > w \circ \nabla w^{\FL} + w^{\FL} + \epsilon \}\bigr).
    \]
    It follows that there is $i_0 \in \N$ such that for all $i \geq i_0$,
    \[
        \alpha * \gamma\bigl(\{v_i \circ \nabla w_i^{\FL} + v_i^{\FL} > w_i \circ \nabla w_i^{\FL} + w_i^{\FL} + \epsilon \}\bigr) \geq 2\delta.
    \]
    Since $\alpha_i * \gamma \to \alpha * \gamma$ in total variation, we can, by possibly further increasing $i_0$, ensure that ${\lVert\alpha * \gamma-\alpha_i * \gamma\rVert_{TV} < \delta}$ for $i \geq i_0$. Thus, for all $i \geq i_0$
    \[
        \alpha_i * \gamma\bigl(\{v_i \circ \nabla w_i^{\FL} + v_i^{\FL} > w_i \circ \nabla w_i^{\FL} + w_i^{\FL} + \epsilon\}\bigr) \geq \delta
    \]
    and consequently, by \cref{lem:strict_descent}, $\Ecal_{\pi'}(w_i) \leq \Ecal_{\pi'}(v_i) - \epsilon\delta$.
\end{proof}

Before turning to the convergence proof, we first require a tightness property of the Martingale Sinkhorn iterates. To this end, we note that the assumption
\[
    \sup_{i \in \N} \int \!\left(\int v_i(y) \, \pi'_x(\rmd y) - v_i^C(x)\right)\mu_0(\rmd x) < \infty,
\]
is naturally satisfied in our setting, since the algorithm decreases the objective at every step. Although it provides only a global control on the objective functional, this condition already enforces local boundedness of the iterates on $I_{\mu_1}$.

\begin{lemme}[Tightness] \label{lem:tightness}
    Let $\mu_0, \mu_1 \in \mathcal{P}_1(\R^d)$ be in convex order and irreducible, and let $(f_n)_{n\in\mathbb{N}}$, $(g_n)_{n\in\mathbb{N}}$ be sequences of l.s.c.~convex functions such that, for all $n\in\mathbb{N}$,
    \[
        I_{\mu_1} \subseteq \operatorname{dom}(f_n), \quad f_n \geq g_n \geq 0, \quad g_n(\bar\mu_0) = 0,
    \]
    and
    \[
        \sup_{n \in \mathbb{N}}\!\int \left( \int f_n(y)\,\pi'_x(\mathrm{d}y) - g_n(x) \right)\mu_0(\mathrm{d}x) < \infty.
    \]
    Then $(f_n)_{n\in\mathbb{N}}$ is locally bounded on $I_{\mu_1}$.
\end{lemme}

\begin{proof} We establish the claim in three steps: \medskip

\emph{Claim 1}: There exist smooth $c\colon \mathbb{R}^d \to\mathbb{R}^+$ and a kernel $(\eta_x)_{x}$ such that, for $\mu_0$-a.e.~$x$,
\begin{align*}
    \epsilon(x) := \eta_x(\R^d) \in \left(0,\tfrac{1}{2}\right], \qquad \eta_x \leq \bigl(1-\epsilon(x)\bigr)\pi'_x, \qquad \overline{\eta}_x = \bigl(1+c(x)\bigr)\bar\mu_0 - c(x) x.
\end{align*}

    Let $\delta \in (0,1)$, $x_1,\dots,x_m \in \supp(\mu_1)$ and $\pi' \in \Cpl_M(\mu_0,\mu_1)$ as in \Cref{prop:existence_super-spreader}. Then, for all $x \in I_{\mu_1}$, the choice 
    \[
        c(x) := \frac{\delta}{1 + \lvert \bar\mu_0 - x\rvert^2}
    \]
    ensures $\bigl(1+c(x)\bigr)\bar\mu_0 - c(x) x \in \bar B_\delta(\bar\mu_0)$. For $i \in \{1,\dots,m\}$, denote $\beta_x^{(i)}$ the barycenter of the restricted measure $\pi'_x(\boldsymbol{\cdot}\,|\,B_\delta(x_i))$. Then, for $\mu_0$-a.e.~$x$, $\bar B_\delta(\bar\mu_0) \subset \interior(\co(\beta_x^{(1)},\dots,\beta_x^{(m)}))$, hence there exists $\alpha_x \in (0,1)^m$ such that 
    \[
        \bigl(1+c(x)\bigr)\bar\mu_0 - c(x) x = \sum_{i=1}^m \alpha_x^{(i)} \beta_x^{(i)}.
    \]
    By \cref{app_B:lem:existence_c1_convex_coeff}, we can choose $\{\alpha_x\}_{x}$ so that $x \mapsto \alpha_x$ is measurable, since each $x \mapsto \beta_x^{(i)}$ is measurable. Consequently, the sub-probability measures
    \[
        \eta_x := \epsilon(x) \sum_{i=1}^m \alpha_x^{(i)} \pi'_x(\boldsymbol{\cdot}\,|\,B_\delta(x_i)), \qquad \epsilon(x) := \frac{\min_k \pi'_x(B_\delta(x_k))}{1+\min_k \pi'_x(B_\delta(x_k))} \in \left(0,\tfrac{1}{2}\right]
    \]
    have barycenter $\bar\eta_x = \bigl(1+c(x)\bigr)\bar\mu_0 - c(x)x$ and satisfy
    \begin{align*}
        \eta_x &\leq \sum_{i=1}^m \frac{\pi'_x(\boldsymbol{\cdot} \cap B_\delta(x_i))}{1+\min_k \pi'_x(B_\delta(x_k))} = (1-\epsilon(x))\sum_{i=1}^m \pi'_x(\boldsymbol{\cdot} \cap B_\delta(x_i)) \leq (1-\epsilon(x))\pi'_x.
    \end{align*}
    On the null set $\big\{x\colon \pi'_x(B_\delta(x_i))=0 \text{ for some }i\big\}$, we set $c(x),\epsilon(x),\eta_x := 0$. \medskip
    
    \emph{Claim 2:} $\displaystyle \int \!\left(\int f_n(y) \, \pi'_x(\rmd y) - g_n(x)\right)\mu_0(\rmd x) \geq \int c(x) \epsilon(x) \int f_n(y)\,\pi'_x(\rmd y) \, \mu_0(\rmd x)$. \medskip
    
    For $\mu_0$-a.e.~$x$, the kernel $\xi := \bigl(c(x)\epsilon(x)\pi'_{x} + \eta_x\bigr)_{x \in I_{\mu_1}}$ satisfies
    \begin{align*}
        &\xi_x \leq \epsilon(x)\pi'_{x} + \eta_x \leq \pi'_x, \qquad \overline{\xi_x} = \frac{c(x)\epsilon(x) \cdot x + \epsilon(x) \cdot \bigl((1+c(x))\bar\mu_0 - c(x)x\bigr)}{c(x)\epsilon(x) + \epsilon(x)} = \bar\mu_0,
    \end{align*}
    hence $\rho_x := \pi'_x - \xi_x$ has mass $\rho_x(\R^d) = 1-\bigl(1+c(x)\bigr)\epsilon(x)$ and barycenter 
    \[
        \overline{\rho_x} = \frac{x-\bar\mu_0\bigl(1+c(x)\bigr)\epsilon(x)}{1-\bigl(1+c(x)\bigr)\epsilon(x)}.
    \]
    Moreover, Jensen's inequality yields $\int g_n \,\mathrm{d}\rho_x \geq \rho_x\!\left(\R^d\right)g_n\!\left(\overline{\rho_x}\right)$ and so
    \begin{align*}
        \int f_n \, \mathrm{d}\pi'_x &= \int f_n \, \mathrm{d}\xi_x + \int f_n \, \mathrm{d}\rho_x \geq \int f_n \, \mathrm{d}\xi_x + \int g_n \, \mathrm{d}\rho_x \\
        &\geq \int f_n \, \mathrm{d}\xi_x + \rho_x\!\left(\R^d\right)g_n\!\left(\overline{\rho_x}\right) + \left(1 -  \rho_x\!\left(\R^d\right)\right)g_n(\bar\mu_0) \\
        & \geq \int f_n \, \mathrm{d}\xi_x + g_n\!\Big(\rho_x\!\left(\R^d\right) \overline{\rho_x}+\left(1-\rho_x\!\left(\R^d\right)\right)\bar\mu_0\Big) \\
        &\geq c(x)\epsilon(x)\!\int f_n \, \mathrm{d}\pi'_x + g_n\!\left(x\right),
    \end{align*}
    since $g_n(\bar\mu_0) = 0$, $f_n \geq g_n\geq0$ and $g_n$ are convex by assumption. Therefore, 
    \[
        \int \!\left(\int f_n(y)\, \pi'_x(\rmd y) - g_n(x) \right)\mu_0(\rmd x) \geq \int c(x)\epsilon(x)\!\int f_n \, \mathrm{d}\pi'_x \,\mu_0(\rmd x).
    \]

    \emph{Claim 3:} $(f_n)_{n \in \N}$ is locally bounded on $I_{\mu_1}$. \medskip
    
    Assume the contrary, i.e., there exist $y_0 \in I_{\mu_1}$, a subsequence $\left(f_{n_k}\right)_{k \in \N}$ and $y_k \in B_{1/k}(y_0)$ such that ${f_{n_k}(y_k) > k}$ for all $k \in \N$. Furthermore, let $\delta > 0$ such that $\bar B_\delta(y_0) \subset I_{\mu_1}$. Given that $y_k \to y_0$, we can assume $y_k \in B_\delta(y_0)$ for all $k \in \N$ without loss of generality.
    
    Fix $k \in \N$. As $f_{n_k}$ is convex, we may find $s_{k} \in \partial f_{n_k}(y_{k})$ such that
    \[
        f_{n_k}(y) \geq f_{n_k}\!\left(y_{k}\right) + s_{k} \cdot \left(y-y_{k}\right) \qquad (\forall\,y \in I_{\mu_1})
    \]
    and since $f_{n_k}$ is non-negative, we obtain, for $\mu_0$-a.e.~$x$, 
    \begin{equation*}
        \int f_{n_k}(y) \, \pi'_x(\rmd y) \geq f_{n_k}\!\left(y_{k}\right) \pi'_x(A) + \int_A s_{k}\cdot\left(y-y_{k}\right) \, \pi'_x(\rmd y)
    \end{equation*}
    for all measurable sets $A$. If $|s_{k}| = 0$, then
    \[
        \int f_{n_k}(y) \, \pi'_x(\rmd y) \geq f_{n_k}\!\left(y_{k}\right) \pi'_x(A),
    \]
    so we only need to treat the case $|s_{k}| \neq 0$. Set $u_k := \frac{s_k}{|s_{k}|} \in \mathcal{S}^{d-1}$ and define for $u \in \mathcal{S}^{d-1}$, ${\kappa \geq 0}$ and $y' \in I_{\mu_1}$,
    \[
        H^+_{u,\kappa}(y') := \{y \in \Rd\colon\, u \cdot (y - y') > \kappa\}.
    \]
    By assumption, $H_{u, \delta}^+(y_0) \neq \emptyset$ for all $u \in \mathcal{S}^{d-1}$ and $|y_k - y_0| \leq \delta$, hence ${H^+_{u_k,\delta}(y_0) \subset H^+_{u_k,0}(y_k)}$. Consequently,
    \[
        \int f_{n_k}(y) \, \pi'_x(\rmd y) \geq f_{n_k}\!\left(y_{k}\right) \pi'_x\!\left(H^+_{u_k,\delta}(y_0)\right)
    \]
    and thus, in every case,
    \[
        \int f_{n_k}(y) \, \pi'_x(\rmd y) \geq f_{n_k}\!\left(y_{k}\right) \! \inf_{u \in \mathcal{S}^{d-1}}\!\pi'_x(H^+_{u,\delta}(y_0)).
    \]
    
    Let $h(u,x) := \pi'_x\bigl(H^+_{u,\delta}(y_0)\bigr)$. It remains to show that $m(x) := \inf_{u \in \mathcal{S}^{d-1}} h(u,x) > 0$ for $\mu_0$-a.e.~$x$. Since $(\pi'_x)_{x}$ is a measurable disintegration and 
    \[
        (u,y)\mapsto \mathbbm 1_{H^+_{u,\delta}(y_0)}(y)
    \]
    is jointly measurable, the map $(x,u)\mapsto h(x,u)$ is jointly measurable too. Moreover, for $\mu_0$-a.e.~$x$, the map $u\mapsto h(x,u)$ is l.s.c.~by Fatou’s lemma. By compactness of $\mathcal S^{d-1}$, for $\mu_0$-a.e.~$x$ the infimum defining $m(x)$ is attained, and ${\Gamma(x) := \{u \in \mathcal S^{d-1} : h(x,u) = m(x)\}}$ is non-empty and compact. Therefore, there exists a measurable selection $x \mapsto u_0(x)$ with $m(x) = \pi'_x\bigl(H^+_{u_0(x),\delta}(y_0)\bigr)$ for $\mu_0$-a.e.~$x$. Since 
    \[
        H^+_{u_0(x),\delta}(y_0) \text{ is open}, \qquad  \mu_1\bigl(H^+_{u_0(x),\delta}(y_0)\bigr) > 0,
    \]
    it follows from \cref{prop:existence_super-spreader} that $m(x) > 0$ for $\mu_0$-a.e.~$x$.
    
    Finally, together with \emph{Claim 2}, we obtain
    \begin{align*}
        \sup_{n \in \N} \int \left(\int f_n(y) \, \pi'_x(\rmd y) - g_n(x)\right)\mu_0(\rmd x) &\geq \liminf_{k \to \infty} \int c(x) \epsilon(x)\int f_{n_k}(y)\,\pi'_x(\rmd y) \, \mu_0(\rmd x) \\
        &\geq  \liminf_{k \to \infty} f_{n_k}(y_k) \int c(x) \epsilon(x) m(x) \, \mu_0(\rmd x) = \infty.
    \end{align*}
    Therefore $(f_n)_{n \in \N}$ is locally bounded.
\end{proof}

Having established both a strict descent property for the algorithm and the tightness of its iterates, we now turn to the convergence of the Martingale Sinkhorn Algorithm.

\begin{theo}[Limit points of the Martingale Sinkhorn Algorithm] \label{thm:limit_points_MS}
    Let Assumptions~\ref{aspt:irreducible}, \ref{aspt:integrable}, \ref{aspt:proper} hold and denote by $(v_i,\alpha_i)_{i \in \N}$ the iterates of \cref{alg:MSinkhorn}. For each $i \in \N$, there exists an affine map $\ell_i$ such that $\hat v_i := v_i + \ell_i$ satisfies $\hat v_i^C \ge 0$ and $\hat v_i^C(\bar\mu_0) = 0$. Set $\hat \alpha_i := (\nabla \hat v_i^C)_\#\mu_0$. Then:
    \begin{enumerate}
        \item The sequence $(\hat v_i)_{i\in\N}$ admits at least one accumulation point on $I_{\mu_1}$ with respect to epi-convergence. Moreover, $(\hat \alpha_i)_{i\in\N}$ is tight in $\Pcal(\Rd)$. \label{en:thm:limit_points_MS:1}
        
        \item Every epi-accumulation point $\hat v$ of $(\hat v_i)_{i\in\N}$ on $I_{\mu_1}$ is a Bass potential, and if $(i_k)_{k\in\N}$ is any subsequence such that $\hat v_{i_k} \to \hat v$ in epi-convergence on $I_{\mu_1}$ then $\hat \alpha_{i_k} \to \hat\alpha$ weakly, where $\hat\alpha$ denotes the Bass measure associated with $\hat v$. \label{en:thm:limit_points_MS:2}
    \end{enumerate}
\end{theo}

\begin{proof}
    \emph{Claim \ref{en:thm:limit_points_MS:1}.}
    Let $v_0 \in L^1(\mu_1)$ l.s.c.~convex such that $v_0^{\FL}*\gamma$ is proper. By \cref{lem:obj_props} and \cref{lem:strict_descent}, for all $i \in \N$,
    \[
        \infty > \Ecal_{\pi'}(v_0) \geq \Ecal_{\pi'}(v_i) \geq \Ecal_{\pi'}(v_{i + 1})\geq 0.
    \]
    For each $i\in\N$, choose an affine map $\ell_i$ as in \cref{lem:ext_obj_props} and set $\hat v_i:=v_i+\ell_i$. Then $\hat v_i^C\ge0$ and $\hat v_i^C(\bar\mu_0)=0$, and by affine invariance, 
    \[
        \sup_{i \in \N} \int \!\left(\int \hat v_i\,\mathrm{d}\pi'_x  - \hat v_i^C(x) \right) \mu_0(\rmd x) < \infty.
    \]
    By \cref{lem:tightness}, the sequence $(\hat v_i)_{i \in \N}$ is locally bounded on $I_{\mu_1}$. By \cref{lem:properties_epi_convergence} together with the subsequent remarks, $(\hat v_i)_{i \in \N}$ admits an epi-accumulation point on $I_{\mu_1}$, which is proper, l.s.c.\ and convex. Moreover, for every $i \in \N$ we have $0 \leq \hat v_i^C \leq \hat v_i$ on $I_{\mu_1}$, and hence $(\hat v_i^C)_{i \in \N}$ is locally bounded on $I_{\mu_1}$. In particular, for every compact $K\subset I_{\mu_1}$ there exists $L_K<\infty$ such that $\partial \hat v_{i}^{C}(x) \subset B_{L_K}(0)$ for all $x \in K$ and all $i \in \N$, and hence $(\hat\alpha_{i})_{j \in \N}$ is tight in $\Pcal(\R^d)$.
    \medskip
    
    \emph{Claim \ref{en:thm:limit_points_MS:2}.}
    Let $(\hat v_{i_j})_{j \in \N}$ be a subsequence that epi-converges to a proper, l.s.c.~convex function $\hat v$ on $I_{\mu_1}$. Since $\hat v_{i_j} \to \hat v$ in epi-convergence, we have $\hat v_{i_j}^{\FL} \to \hat v^{\FL}$ in epi-convergence (see \cref{lem:properties_epi_convergence}). Since $(\nabla \hat v_{i_j}^{\FL})_\# \hat \alpha_{i_j}* \gamma = \mu_1$, \cref{lem:limit_potential} yields ${(\nabla \hat v^{\FL})_\# \hat\alpha * \gamma = \mu_1}$ for every accumulation point $\hat\alpha$ of $(\hat \alpha_{i_j})_{j\in\mathbb N}$. Moreover, by \cref{lem:exp.bound} there exist $c, R > 0$ such that for all $|x| \geq R$,
    \[
        \sup_{j \in \N} |\hat v_{i_j}^{\FL}(x)| \leq c\exp\!\left(\tfrac2{3+p}|x|^2 \right) \in L^1(\gamma).
    \]
    Dominated convergence then implies $\hat v_{i_j}^{\FL} * \gamma \to \hat v^{\FL} * \gamma$ pointwise, and thus ${\hat v^C_{i_j} \to \hat v^C}$ in epi-convergence.
    Because $\hat v^C$ is differentiable on $I_{\mu_1} \subset \operatorname{int}(\operatorname{dom}(\hat v^C))$ (see \cref{lem:ext_obj_props} and the proof thereof), we obtain $\nabla \hat v^C_{i_j} \to \nabla \hat v^C$ pointwise on $I_{\mu_1}$ and therefore
    \[
        (\nabla \hat v^C)_\# \mu_0 = \lim_{j \to \infty} \hat \alpha_{i_j} = \hat \alpha.
    \]
    Likewise, by \cref{lem:limit_potential} (and passing to a further subsequence if necessary),
    \[
        \lim_{j \to \infty} \hat v_{i_j+1} = \hat w_{+1} \ \text{in epi-convergence}, \quad
        (\nabla \hat w_{+1}^{\FL})_\# \hat \alpha * \gamma = \mu_1.
    \]
    Thus $\hat w_{+1}$ is the Martingale Sinkhorn iterate following $\hat v$. If $\hat v$ and $\hat w_{+1}$ are not equal up to an additive constant, then by \cref{cor:the_universal_gap} there exist $\varepsilon > 0$ and $j_0 \in \N$ such that for all $j \ge j_0$,
    \[
        \Ecal_{\pi'}(\hat v_{i_j}) - \varepsilon \geq \Ecal_{\pi'}(\hat v_{i_j + 1})\geq \Ecal_{\pi'}(\hat v) \geq 0,
    \]
    hence $\Ecal_{\pi'}(\hat v_{i_j}) - k\,\varepsilon \geq \Ecal_{\pi'}(\hat v_{i_j+k}) \geq \Ecal_{\pi'}(\hat v) \geq 0$ for all $k \in \N$, a contradiction. Therefore $\hat v = \hat w_{+1} + c$ for some $c \in \R$, and so $(\nabla \hat v^C)_\# \mu_0 = \hat \alpha$ and $(\nabla \hat v^\FL)_\# (\hat\alpha*\gamma) = (\nabla \hat w_{+1}^\FL)_\# (\hat\alpha*\gamma) = \mu_1$. Hence $\hat v$ is a Bass potential. 
\end{proof}
\medskip

The previous result shows that the Martingale Sinkhorn algorithm produces Bass potentials as limit points. We next address uniqueness and the relation to stretched Brownian motion. The following lemma shows that a Bass coupling determines its potential uniquely, up to an affine function.

\begin{lemme} \label{lem:bass_affine_difference}
    Let Assumption~\ref{aspt:irreducible} hold, and let $v,w$ be two Bass potentials whose associated Bass couplings coincide. Then $v-w$ is an affine function on $I_{\mu_1}$.
\end{lemme}

\begin{proof}
    If $v,w$ induce the same Bass coupling $\pi \in \Cpl_M(\mu_0,\mu_1)$, then, for $\mu_0$-a.e.~$x$,
    \[
        \bigl(\nabla v^\FL(\cdot +\nabla v^C(x))\bigr)_\#\gamma = \bigl(\nabla w^\FL(\cdot +\nabla v^C(x))\bigr)_\#\gamma.
    \]
    For fixed $x$ the functions ${z\mapsto v^\FL(z +\nabla v^C(x))}$ and ${z\mapsto w^\FL(z +\nabla v^C(x))}$ are proper, l.s.c.\ convex potentials, and their gradients push $\gamma$ to the same measure $\pi_x$. By the Brenier--McCann uniqueness theorem, they differ only by an additive constant $b(x) \in \R$. Taking convex conjugates yields, for $\mu_0$-a.e.~$x$ and all $y \in \R^d$,
    \[
        v(y) - y\, \nabla v^C(x) = w(y) - y\, \nabla w^C(x) + b(x),
    \]
    i.e., $v(y) - w(y) = y\cdot\bigl(\nabla v^C(x) - \nabla w^C(x)\bigr) + b(x)$. We conclude that $v-w$ is an affine function on $I_{\mu_1}$, as claimed.
\end{proof}

\begin{prop}[Stretched BM and Bass potential attaining~\eqref{eq:dmbb}] \label{prop:some_basspotential_attains_dmbb} Let Assumption~\ref{aspt:irreducible} hold and recall that ${\pi^{\rm SBM} \in \Cpl_M(\mu_0,\mu_1)}$ denotes the unique solution to~\eqref{eq:mbb}. Then:
\begin{enumerate}
    \item There exists a Bass potential $\psi_\infty$ which attains the dual problem~\eqref{eq:dmbb}, and whose associated Bass coupling is $\pi^{\rm SBM}$; \label{en:prop:some_basspotential_attains_dmbb:1}
    \item If $v$ is a Bass potential with $v \in L^1(\mu_1)$, then $v$ attains the dual problem~\eqref{eq:dmbb}, and the associated Bass martingale coupling is $\pi^{\rm SBM}$. \label{en:prop:some_basspotential_attains_dmbb:2}
\end{enumerate}
\end{prop}

\begin{proof}
\emph{Claim \ref{en:prop:some_basspotential_attains_dmbb:1}.}
    Let $(\psi_n)_{n\in\N}$ be a minimising sequence for the dual problem \eqref{eq:dmbb}. Note that $\pi^{\rm SBM}$ is a super-spreader in the sense of \cref{prop:existence_super-spreader}. Therefore, after affine normalisation and by tightness (see \cref{lem:tightness}), the same arguments as in the proof of \cref{thm:limit_points_MS} \eqref{en:thm:limit_points_MS:1} yields a proper l.s.c.~convex epi-limit potential $\psi_\infty$ with
    \[
        \int \MCov(\pi^{\rm SBM}_x,\gamma)\,\mathrm{d}\mu_0(x) = \liminf_{n \to \infty} \Ecal(\psi_n) = \liminf_{n \to \infty} \Ecal_{\pi^{\rm SBM}}(\psi_n) \geq \Ecal_{\pi^{\rm SBM}}(\psi_\infty).
    \]
    Since $\int \MCov(\pi_x,\gamma) \,\mu_0(\mathrm{d}x) \leq \Ecal_{\pi}(\psi)$ for all $\pi \in \Cpl_M(\mu_0,\mu_1)$ and all convex $\psi$, we obtain
    \[
        \int \MCov(\pi^{\rm SBM}_x,\gamma)\,\mathrm{d}\mu_0(x) \leq \Ecal_{\pi^{\rm SBM}}(\psi_\infty) \leq \int \MCov(\pi^{\rm SBM}_x,\gamma)\,\mathrm{d}\mu_0(x).
    \]
    Hence $\MCov(\pi^{\rm SBM}_x,\gamma) = \int \psi_\infty\,\mathrm{d}\pi^{\rm SBM}_x - \psi_\infty^C(x)$ for $\mu_0$-a.e.~$x$. By the Fenchel--Legendre duality,
    \begin{align*}
        \MCov(\pi^{\rm SBM}_x,\gamma) &= \int \psi_\infty\,\mathrm{d}\pi^{\rm SBM}_x - x \nabla\psi_\infty^C(x) + (\psi_\infty^\star*\gamma) \circ \nabla \psi_\infty^C(x) \\
        &= \int \left(\psi_\infty(y) - y\,\nabla\psi_\infty^C(x) \right)\,\pi^{\rm SBM}_x(\rmd y) + \int \psi_\infty^\star\!\left(z+\nabla \psi_\infty^C(x)\right)\,\gamma(\rmd z).
    \end{align*}
    Since $\left(\psi_\infty - \operatorname{id}a\right)^\star(z) = \psi_\infty^\star\!\left(z+a\right)$ for every $z, a \in \Rd$, it follows that ${y\mapsto\psi_\infty(y) - y\,\nabla\psi_\infty^C(x)}$ is the dual maximiser of $\MCov(\pi^{\rm SBM}_x,\gamma)$ for $\mu_0$-a.e.~$x$, that is, the Brenier--McCann potential satisfying
    \[
        \left(\nabla \psi_\infty^\star\!\left(\,\cdot\,+\nabla \psi_\infty^C(x)\right)\right)_\# \gamma = \pi^{\rm SBM}_x.
    \]
    Hence $\psi_\infty$ is a Bass potential, and its corresponding Bass coupling is $\pi^{\rm SBM}$.
    \medskip

    \emph{Claim \ref{en:prop:some_basspotential_attains_dmbb:2}.}
    Let $v$ be a Bass potential and denote $\pi \in \Cpl_M(\mu_0,\mu_1)$ the associated Bass coupling. By definition, ${\pi_x = \bigl(\nabla v^\FL(\cdot +\nabla v^C(x))\bigr)_\#\gamma}$ for $\mu_0$-a.e.~$x$. Since ${(v - \operatorname{id}a)^\star(z) = v^\star(z+a)}$ for every $z, a \in \Rd$, it follows that the potential $y \mapsto v(y) - y\,\nabla v^C(x)$ is a dual optimiser of $\MCov(\pi_x,\gamma)$ for $\mu_0$-a.e.~$x$. Hence
    \[
        \int \MCov(\pi_x,\gamma) \,\mu_0(\mathrm{d}x) = \int \! \left(\int v\,\rmd\pi_x - x\,\nabla v^C(x) + (v^\FL*\gamma) \bigl(\nabla v^C(x)\bigr)\right)\mu_0(\mathrm{d}x) = \Ecal_{\pi}(v).
    \]
    If in addition $v \in L^1(\mu_1)$, then by strong duality (see \cref{app:thm:strong_duality_mBB}) we obtain
    \[
        \int \MCov(\pi_x,\gamma) \,\mu_0(\rmd x)= \Ecal_{\pi}(v) = \Ecal(v) \geq \inf_{\substack{\psi \in L^1(\mu_1),\\ \text{convex}}} \Ecal(v) = \sup_{\pi \in \Cpl_M(\mu_0,\mu_1)} \int \MCov(\pi_x,\gamma)\,\mu_0(\mathrm{d}x),
    \]
    Thus $v$ attains the dual problem \eqref{eq:dmbb}, and $\pi$ solves the primal problem \eqref{eq:mbb}. By uniqueness, $\pi = \pi^{\rm SBM}$.
\end{proof}

Under the normalisation in \cref{thm:limit_points_MS} and by the above results, \cref{alg:MSinkhorn} has at most one epi-accumulation point in $L^1(\mu_1)$. Moreover, convergence holds if the Bass coupling is unique, or equivalently (by \cref{lem:bass_affine_difference}) if the Bass potential is unique up to translation by an affine function.

\begin{cor}\label{cor:uniquenss_algoconverges}
    In the setup of \cref{thm:limit_points_MS}, assume that $\pi^{\rm SBM}$ is the unique Bass coupling. Then the iterates $(\hat v_i,\hat\alpha_i)_{i\in\N}$ epi-converge to the (normalised) Bass potential and its Bass measure.
\end{cor}

\begin{proof}
By \cref{thm:limit_points_MS}, every subsequence of $(\hat v_i,\hat\alpha_i)_{i\in\N}$ admits a subsequence epi-converging to a Bass potential and its Bass measure. If $\pi^{\rm SBM}$ is the unique Bass coupling, then the Bass measure is unique and the Bass potential is unique up to affine shifts; under the normalisation of \cref{thm:limit_points_MS} the limit is therefore unique. Hence the whole sequence epi-converges to the (normalised) Bass potential and its Bass measure.
\end{proof}

In dimension one, the following result shows that the Bass potential itself is unique up to affine normalisation and that the associated coupling is the stretched Brownian motion, $\pi^{\rm SBM}$. 

\begin{prop}[Uniqueness of the Bass potential on the line] \label{prop:unique_basspotential_R1}
    Let $\mu_0, \mu_1 \in \Pcal_p(\R)$ for some $p > 1$ be in convex order and irreducible. Then the Bass potential associated with $(\mu_0,\mu_1)$ is unique up to affine normalisation, and its associated Bass coupling is $\pi^{\rm SBM}$.
\end{prop}

\begin{proof}
    By \cref{prop:some_basspotential_attains_dmbb} there exists a Bass potential $\psi_\infty$ which attains~\eqref{eq:dmbb}, and its corresponding Bass coupling is $\pi^{\rm SBM}$, the unique solution to~\eqref{eq:mbb}. It remains to show that $\psi_\infty$ is unique up to affine normalisation.
    
    Let $\hat v$ be another Bass potential, and denote by $\hat\pi \in \Cpl_M(\mu_0,\mu_1)$ its associated Bass coupling. Let $(K_n)_{n\in\N}$ be an increasing sequence of compact sets with $K_n \uparrow I_{\mu_1}$ and $\mu_0(K_n)>0$ for all $n \in \N$. For each $n \in \N$ set
    \[
        \mu_0^{(n)} := \frac{\mu_0(\cdot \cap K_n)}{\mu_0(K_n)}, \quad \mu_1^{(n)} := \int \hat\pi_x\,\mu_0^{(n)}(\rmd x).
    \]
    Then $\mu_0^{(n)} \to \mu_0$ and $\mu_1^{(n)} \to \mu_1$ in total variation. Fix $n \in \N$. We have $\mu_0^{(n)} \leq_{\rm cvx} \mu_1^{(n)}$, and since $\hat \pi$ is a super-spreader in the sense of \cref{prop:existence_super-spreader}, the pair $(\mu_0^{(n)},\mu_1^{(n)})$ is irreducible as well. By construction, $\hat \pi$ is the unique solution to
    \[
        \mathbf{P}^{(n)}_{\mu_0,\mu_1}:= \sup_{\pi \in \Cpl_M(\mu_0^{(n)},\mu_1^{(n)})} \int \MCov(\hat\pi_x,\gamma)\,\mu_0^{(n)}(\rmd x),
    \]
    i.e., to the martingale Benamou--Brenier problem for the pair $\bigl(\mu_0^{(n)}, \mu_1^{(n)}\bigr)$. By \cite[Theorem 2.8]{beiglbock2023stability}, we have $\mathbf{P}^{(n)}_{\mu_0,\mu_1} \to \mathbf{P}_{\mu_0,\mu_1}$ and any accumulation point of the primal optimisers for $\bigl(\mathbf{P}^{(n)}_{\mu_0,\mu_1}\bigr)_{n\in \N}$ is a maximiser of $\mathbf{P}_{\mu_0,\mu_1}$. It follows that $\hat \pi = \pi^{\rm SBM}$. By \cref{lem:bass_affine_difference}, we conclude that $\hat v - \psi_\infty$ is an affine function on $I_{\mu_1}$.
\end{proof}

In dimension one the Bass coupling is unique, and \cref{cor:uniquenss_algoconverges} therefore yields convergence in \cref{thm:limit_points_MS}. More generally, the same argument would give uniqueness of the Bass coupling (and hence of the normalised Bass potential) in any dimension, provided one has an appropriate stability result for weak martingale optimal transport. We expect that \cite[Theorem~2.8]{beiglbock2023stability} extends to $\R^d$ under suitable assumptions, and defer this to future work.

\section{Implementation and examples} \label{sec:implementation_and_examples}

\begin{figure}[t]
  \centering
  \includegraphics[width=0.5\textwidth]{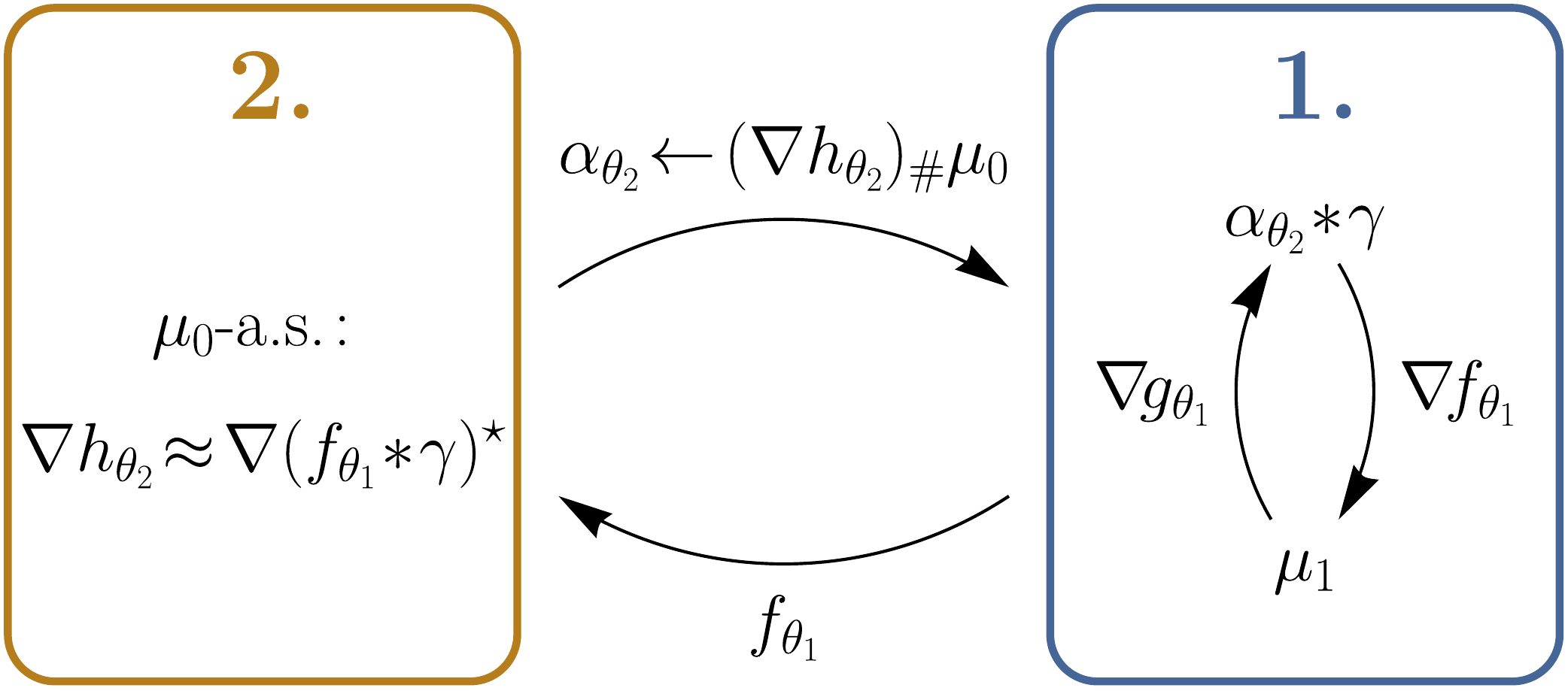}
  \caption{One iteration of the Martingale Sinkhorn scheme. 1.~Optimal transport step (right): for the current latent proposal $\alpha_{\theta_2}$, learn dual OT potentials $f_{\theta_1}, g_{\theta_1}$ for the transport problem between $\alpha_{\theta_2} * \gamma$ and $\mu_1$. 2.~Generator step (left): on samples from $\mu_0$, train $h_{\theta_2}$ so that $\mu_0$-a.s.\ $\nabla h_{\theta_2} \approx \nabla(f_{\theta_1} * \gamma)^\star$.}
  \label{fig:msinkhorn-iteration-scheme}
\end{figure}

We briefly describe the numerical implementation of the Martingale Sinkhorn algorithm, followed by examples for selected pairs of distributions $\mu_0, \mu_1 \in \Pcal_2(\mathbb{R}^2)$. Given samples $(x_i)_{i=1}^{N_0} \sim \mu_0$ and $(y_j)_{j=1}^{N_1} \sim \mu_1$, we work with the corresponding empirical measures and use a neural network approximation of the Bass system depicted in \cref{fig:MS_algorithm}. In the spirit of existing neural OT approaches (see e.g.~\cite{makkuva2020optimal,amos2022amortizing}), we decompose one iteration of the Martingale Sinkhorn scheme into two subroutines (see \cref{fig:msinkhorn-iteration-scheme}), which are applied in an alternating fashion: An \emph{optimal transport step}, where the dual potentials are updated for the current generator, and a \emph{generator step}, where the generator is updated to match the dual potentials.

\subsection*{1.~Optimal transport step:}
Given a proposal $\alpha_{\theta_2} \in \Pcal_2(\R^d)$, this step solves the optimal transport problem for the pair $(\alpha_{\theta_2} * \gamma,\mu_1)$. This amounts to finding the optimal dual potential $\psi$ that achieves
\begin{equation} \label{eq:MCov_loss}
    \MCov(\alpha_{\theta_2}*\gamma,\mu_1) = \inf_{\psi \text{ convex, l.s.c.}} \int \psi^\star \,\mathrm{d}(\alpha_{\theta_2}*\gamma) + \int \psi \,\mathrm{d}\mu_1.
\end{equation}
Since $\alpha_{\theta_2},\mu_1 \in \Pcal_2(\R^d)$, we equivalently solve
\begin{equation} \label{eq:W2_loss}
    \Wcal_2^2(\alpha_{\theta_2}*\gamma,\mu_1) = \sup_{\varphi \in L^1(\mu_1)} \int q_2\square(-\varphi)\,\mathrm{d}(\alpha_{\theta_2}*\gamma) + \int \varphi\,\mathrm{d}\mu_1,
\end{equation}
where
\[
    \bigl(\tfrac{1}{2}|\cdot|^2\bigr)\square(-\varphi)(x) := \inf_{y \in \R^d} \tfrac{1}{2}|x-y|^2 - \varphi(y)
\]
is the infimal convolution of $\tfrac{1}{2}|\cdot|^2$ and $-\varphi$. In particular, the optimal potentials $\psi$ and $\varphi$ for \eqref{eq:MCov_loss} and \eqref{eq:W2_loss}, respectively, satisfy 
\[
    \psi = \tfrac{1}{2}|\cdot|^2-\varphi, \qquad \psi^\star = \tfrac{1}{2}|\cdot|^2 - \bigl(\tfrac{1}{2}|\cdot|^2\bigr)\square(-\varphi).
\]

For the implementation of the numerical scheme we restrict to potentials with quadratic growth at infinity. More precisely, we parameterise the candidate potential $\psi$ as $\psi \approx g_{\theta_1} := \tfrac{1}{2}|\cdot|^2 - \tilde g_{\theta_1}$ where $\tilde g_{\theta_1}$ is an unconstrained multilayer perceptron (MLP) with activation functions that grow at most linearly at infinity (for instance ReLU, ELU or SiLU). Then $\tilde g_{\theta_1}$ has at most linear growth, so $g_{\theta_1}$ retains quadratic growth at infinity, i.e., $g_{\theta_1} \simeq \tfrac{1}{2}|\cdot|^2$. This quadratic-growth class is stable under convex conjugation, and hence $g_{\theta_1}^\FL \simeq \tfrac{1}{2}|\cdot|^2$ at infinity too. Accordingly, we parameterise $g_{\theta_1}^\FL$ by a function of the same form, i.e., $g_{\theta_1}^\FL \approx f_{\theta_1} := \tfrac{1}{2}|\cdot|^2 - \tilde f_{\theta_1}$ where $\tilde f_{\theta_1}$ is another unconstrained MLP of the same type, so that all potentials appearing in the dual formulations remain in the same quadratic-growth class.

Having fixed the parameterisation $(f_{\theta_1},g_{\theta_1})$, we train $\theta_1$ using the expectile–regularised neural OT (ENOT) method of~\cite{buzun2024expectile}, specialised to the quadratic cost $c(x,y):=\tfrac{1}{2}|x-y|^2$ with $\alpha \gets \alpha_{\theta_2} * \gamma$ and $\beta \gets \mu_1$. We use their \emph{bidirectional training mode} (Algorithm~1 in~\cite{buzun2024expectile}), in which the roles of $\tilde f_{\theta_1}$, $\tilde g_{\theta_1}$ and
of the two measures in the ENOT loss are alternately swapped at each optimisation step. An efficient implementation of this algorithm, based on empirical measures, is available in the \texttt{OTT} library by \cite{cuturi2022ott_jax}, which we rely upon in our implementation of \cref{alg:MSinkhorn}.

\subsection*{2.~Generator step:}
Given $f_{\theta_1}$, this step computes the new latent proposal distribution
\[
  \alpha_{\theta_2} :\approx \bigl(\nabla (f_{\theta_1}*\gamma)^\star\bigr)_{\#}\mu_0.
\]
To this end we introduce a third potential $h_{\theta_2}$ and require $h_{\theta_2} \approx (f_{\theta_1}*\gamma)^\star$. By construction $f_{\theta_1} \simeq \tfrac12|\cdot|^2$, hence also $f_{\theta_1}*\gamma \simeq \tfrac12|\cdot|^2$ and $(f_{\theta_1}*\gamma)^\star \simeq \tfrac12|\cdot|^2$ at infinity. We therefore parameterise $h_{\theta_2}(x) := \tfrac12|x|^2 - \tilde h_{\theta_2}(x)$ where $\tilde h_{\theta_2}$ is an unconstrained MLP of the same type as above.

To train $\theta_2$ we exploit the Fenchel--Young inequality for $f_{\theta_1}*\gamma$: For all $x,y \in \mathbb R^d$,
\[
  (f_{\theta_1}*\gamma)(y) - (f_{\theta_1}*\gamma)^\star(x) \geq x\,y,
\]
with equality if and only if $y \in \partial (f_{\theta_1}*\gamma)^\star(x)$. Since $(f_{\theta_1}*\gamma)^\star$ is differentiable on
$I_{\mu_1}\subset \operatorname{int}(\operatorname{dom}((f_{\theta_1}*\gamma)^\star)) \neq \emptyset$
(see \cref{lem:ext_obj_props} and the proof thereof), this reduces to the condition $y = \nabla (f_{\theta_1}*\gamma)^\star(x)$. This leads to the Fenchel--Young loss
\begin{equation}\label{eq:FY-loss}
  \mathcal L_{\mathrm{FY}}(\theta_2) = \int (f_{\theta_1}*\gamma)\bigl(\nabla h_{\theta_2}(x)\bigr) - (f_{\theta_1}*\gamma)^\star(x) - x\,\nabla h_{\theta_2}(x)\,\mu_0(\rmd x),
\end{equation}
with $\mathcal L_{\mathrm{FY}}(\theta_2) = 0$ if and only if $\nabla h_{\theta_2}(x) = \nabla (f_{\theta_1}*\gamma)^\star(x)$ for $\mu_0$-a.e.~$x$. In the numerical implementation we replace $\mu_0$ by its empirical measure. The functional $\mathcal L_{\mathrm{FY}}$ is a Fenchel--Young-type loss in the sense of \cite{blondel2020learning}. Closely related amortised conjugacy losses have also been used in the neural OT literature, see for instance \cite{amos2022amortizing,buzun2024expectile}. 

Once $\theta_2$ has been trained we set $\alpha_{\theta_2} := (\nabla h_{\theta_2})_{\#}\mu_0$ in preparation for the next optimal transport step. Since $h_{\theta_2} \simeq \tfrac{1}{2}|\cdot|^2$ at infinity, we conclude by the transport relation
\[
    \int |x|^2\,\alpha_{\theta_2}(\mathrm{d}x) = \int |\nabla h_{\theta_2}|^2\,\mathrm{d}\mu_0 < \infty,
\]
so that each iterate $\alpha_{\theta_2}$ lies in $\Pcal_2(\R^d)$ and the quadratic moment condition is preserved along the scheme.

\medskip

Starting from $v_0(x) := \tfrac{1}{2}\lVert x\rVert_2^2$ so that $\alpha_{\theta_2} = \mu_0$ in the first call of the optimal transport step, alternating between $n_1$ optimal transport steps and $n_2$ generator steps yields the Martingale Sinkhorn algorithm.
\medskip

\begin{remarque}~
\begin{enumerate}
    \item Note that $(f_{\theta_1}*\gamma)(x) = \mathbb{E}_{Z\sim\gamma}\big[f_{\theta_1}(x+Z)\big]$. In our implementation, we approximate this potential in the generator step using a Monte Carlo estimator. This is the most computationally expensive part of the implementation and could be further improved using standard variance reduction techniques. 
    \item OT step is expected to be harder than the generator step and our numerical illustrations confirmed that choosing $n_1 \gg n_2$ proves effective.
    \item After each full iteration of the Martingale Sinkhorn scheme we reset the optimiser states, but warm-start the training by reusing the parameters from the previous iteration. In particular, we do not perform an affine normalisation of the potentials. 
    \item If $\mu_0, \mu_1 \in \Pcal_p(\R^d)$ with $1 < p < 2$, the Martingale Sinkhorn algorithm can be implemented analogously: In the optimal transport step, we again approximate the dual minimisers $\psi$ and $\psi^\star$ in \eqref{eq:MCov_loss}. In this case, we can parameterise the candidate potentials $\psi$, denoted $g_{\theta_1}$, by $g_{\theta_1} := \tfrac{1}{p}|\cdot|^p - \tilde g_{\theta_1}$ where $\tilde g_{\theta_1}$ is as above. Indeed, $g_{\theta_1} \simeq \tfrac{1}{p}|\cdot|^p$ at infinity, hence $g_{\theta_1}^\star \simeq \tfrac{1}{q}|\cdot|^q$ with $q > 1$ such that $\tfrac{1}{p}+\tfrac{1}{q}=1$. Therefore, we can approximate $g_{\theta_1}^\star$ by $f_{\theta_1} := \tfrac{1}{q}|\cdot|^q - \tilde f_{\theta_1}$ with $\tilde f_{\theta_1}$ as above, and proceed analogously.
    Regarding the parameterisation of $h_{\theta_2}$ in the generator step, we note that $f_{\theta_1} * \gamma \simeq \tfrac{1}{q}|\cdot|^q$ at infinity, hence ${(f_{\theta_1} * \gamma)^\star \simeq \tfrac{1}{p}|\cdot|^p}$. Parameterising $h_{\theta_2} := \tfrac{1}{p}|\cdot|^p -\tilde h_{\theta_2}$ with $\tilde h_{\theta_2}$ as above, we obtain ${\alpha_{\theta_2}:= (\nabla h_{\theta_2})_\#\mu_0 \in \Pcal_p(\R^d)}$ throughout all iterations.
\end{enumerate}
\end{remarque}

\subsection{Examples}
In this subsection we present three two-dimensional examples, in the first of which we compute the Bass potential and measure explicitly. In each example, we visualise the learned martingale using three diagnostic plots, each consisting of three heat-maps. In \cref{ex:gaussian_mixture_refinement} and \cref{ex:moons_to_8_gaussians}, the three heat-maps (from top to bottom) compare (i) the source distribution $\mu_0$ with $(\nabla (f_{\theta_1}*\gamma))_\#\alpha_{\theta_2}$ where $\alpha_{\theta_2} = (\nabla h_{\theta_2})_\#\mu_0$, (ii) the intermediate law $\alpha_{\theta_2} * \gamma$ with $(\nabla g_{\theta_1})_\# \mu_1$, and (iii) the target marginal $\mu_1$ with $(\nabla h_{\theta_2})_\# (\alpha_{\theta_2} * \gamma)$. In \cref{ex:uniform_disk_to_circle}, where an explicit expression for the Bass measure is available, the middle panel instead compares the true Bass measure $\alpha$ with $\alpha_{\theta_2}$.

\begin{remarque}
    All computations are performed in \texttt{JAX}. The code to reproduce all experiments and figures is available in the public GitHub repository \href{https://github.com/manuelhasenbichler/martingale-sinkhorn}{\texttt{martingale-sinkhorn}}.
    
    In \cref{ex:gaussian_mixture_refinement,ex:moons_to_8_gaussians} the potentials $f_{\theta_1}$, $g_{\theta_1}$, $h_{\theta_2}$ are parameterised by fully connected neural networks with three hidden layers of width $64$ and SiLU activations, followed by a scalar output layer. In \cref{ex:uniform_disk_to_circle} we use four such hidden layers. All three networks are trained with the Adam optimiser (learning rate $5\times 10^{-4}$). The ENOT hyperparameters are fixed to $\tau = 0.98$ and $\lambda = 0.5$.
    
    For all three examples we construct empirical measures from $20\,000$ independent samples from each marginal. In \cref{ex:gaussian_mixture_refinement} we perform $5$ Martingale Sinkhorn iterations, each consisting of $n_1 = 4000$ optimal transport steps and $n_2 = 200$ generator steps. In \cref{ex:moons_to_8_gaussians,ex:uniform_disk_to_circle} we instead use $10$ Martingale Sinkhorn iterations with the same $(n_1,n_2) = (4000,200)$. During training, $f_{\theta_1} * \gamma$ is approximated by Monte Carlo simulation with $256$ samples in \cref{ex:gaussian_mixture_refinement,ex:moons_to_8_gaussians} and $512$ samples in \cref{ex:uniform_disk_to_circle}.
    
    For all visualisations, we draw $40\,000$ independent samples from each marginal and approximate $f_{\theta_1} * \gamma$ using $512$ Monte Carlo samples. All plots are generated on a validation set.
\end{remarque}
\medskip

\begin{example}[Uniform Disk to Uniform Circle]  \label{ex:uniform_disk_to_circle}
    We start with an example in which case we can explicitly compute the Bass potential and measure. The corresponding diagnostic plots are shown in \cref{fig:ex1-diag}. Let $Y\sim\mu_1$ be uniform on the unit circle $\mathbb S^1$, let $R$ be independent of $Y$ with density $f_R(r)=2r\,\mathbf 1_{[0,1]}(r)$, and set $X:=RY$. Then $X\sim\mu_0$ is uniform on the unit disk and $\mu_0\le_{cvx}\mu_1$, since for convex $\psi\colon \R^2\to \R$, $\psi(RY) \leq (1-R)\psi(0)+R\psi(Y)$, and
    \[
        \E[\psi(X)]\le (1-\E[R])\,\psi(0)+\E[R]\,\E[\psi(Y)] = (1-\E[R])\,\psi(\E[Y])+\E[R]\,\E[\psi(Y)] \le \E[\psi(Y)].
    \]    
    For $v^\star(x)=|x|$ and $Z\sim\Ncal(0,I_2)$,
    \[
        (\nabla v^\star * \gamma)(y)=\E\!\left[\frac{y+Z}{|y+Z|}\right]
        = h(|y|)\frac{y}{|y|}
    \]
    is a rotation-equivariant map $\R^2 \to \R^2$. With $I_\nu$ being the modified Bessel function of the first kind, we have
    \[
        h(r)=\sqrt{\frac{\pi}{8}}\, r\, e^{-\tfrac{r^2}4}\!\left(I_0\!\left(\tfrac{r^2}{4}\right)+I_1\!\left(\tfrac{r^2}{4}\right)\right),
    \]
    see, e.g., \cite{kendall1974pole}. Let $\alpha$ be the law of $h^{-1}(R)\,Y$. Then
    \[
        (\nabla v^\star * \gamma)(h^{-1}(R)\,Y)=R\,Y,
    \]
    so $(\nabla v^\star * \gamma)_\#\alpha=\mu_0$. Moreover,
    \[
        \nabla v^\star(h^{-1}(R)\,Y+Z)=\frac{h^{-1}(R)\,Y+Z}{|h^{-1}(R)\,Y+Z|}\in\mathbb S^1
    \]
    has a rotation-invariant law, i.e., the unique uniform law on $\mathbb S^1$, hence $(\nabla v^\star)_\# \alpha*\gamma = \mu_1$. Finally, using the asymptotic expansion of the modified Bessel functions, we obtain, as $\rho \to \infty$,
    \[
        \mathbb{P}\big[h^{-1}(R) > \rho\big] = 1 - h(\rho)^2 \simeq \rho^{-2}.
    \]
    In particular, its Bass measure $\alpha$ does not admit a finite second moment, although $\mu_0$ and $\mu_1$ are compactly supported. This illustrates that the Bass measure may exhibit substantially heavier tails than its marginals.

    \begin{remarque}\label{rk:alphano1moment}
        Replacing the uniform measure on the disk by a measure placing more mass towards the boundary, for instance
        \[
            f_R(r) = \frac{1}{2}(1-r)^{-\frac{1}{2}}\,\mathbf 1_{[0,1]}(r),
        \]
        we obtain, as $\rho\to\infty$,
        \[
            \mathbb{P}\big[h^{-1}(R) > \rho\big] = \sqrt{1 - h(\rho)} \simeq \frac{1}{\sqrt{2}\rho}.
        \]
        In particular, the associated Bass measure $\alpha$ does not admit a finite first moment, and $\MCov(\mu_0,\alpha) = \infty$.
    \end{remarque}
\end{example}

\medskip

\begin{example}[Gaussian Mixture Refinement] \label{ex:gaussian_mixture_refinement}
    In this example, we start from a three-component $2$-dimensional Gaussian mixture
    \[
    \mu_0=\sum_{i=1}^3 p_i\,\mathcal N_2(\eta_i,\Sigma_i),
    \]
    whose parameters are reported in Table~\ref{tab:ex2-mu0}. The corresponding diagnostic plots are shown in \cref{fig:ex2-diag}. To obtain a target that dominates $\mu_0$ in convex order, we refine each component $i$ into three subcomponents
    $j\in\{1,2,3\}$ with weights $q$ and define
    \[
    \mu_1=\sum_{i=1}^3\sum_{j=1}^3 p_i q_j\,
    \mathcal N_2(\eta_i+s_{ij},\,\Sigma_i+C_{ij}),
    \]
    where the subcomponent shift $s_{ij}\in\R^2$ is chosen to be mean-preserving within each $i$ by setting
    \[
    s_{i3}:=-\frac{q_1 s_{i1}+q_2 s_{i2}}{q_3}.
    \]
    Finally, the additional covariance $C_{ij}$ is parameterized via $(\sigma_{0,ij},\sigma_{1,ij},\rho_{ij})$ as
    \[
    C_{ij}=
    \begin{bmatrix}
    \sigma_{0,ij}^2 & \rho_{ij}\sigma_{0,ij}\sigma_{1,ij}\\
    \rho_{ij}\sigma_{0,ij}\sigma_{1,ij} & \sigma_{1,ij}^2
    \end{bmatrix}.
    \]
    All $(s_{ij},\sigma_{0,ij},\sigma_{1,ij},\rho_{ij})$ are given in Table~\ref{tab:ex2-mu1}.
    
    \begin{table}[h]
        \centering
        \renewcommand{\arraystretch}{1.25}
        \setlength{\tabcolsep}{7pt}
        \begin{tabular}{cccc}
            \toprule
            $i$ & $p_i$ & $\eta_i^\top$ & $\Sigma_i$\\
            \midrule
            1 & 0.5 & $[0,\,0]$ &
            $\begin{bmatrix}1 & 0.4\\ 0.4 & 0.7\end{bmatrix}$\\[4pt]
            2 & 0.3 & $[2.5,\,1]$ &
            $\begin{bmatrix}0.8 & -0.2\\ -0.2 & 0.5\end{bmatrix}$\\[4pt]
            3 & 0.2 & $[-2,\,2.5]$ &
            $\begin{bmatrix}0.6 & 0.3\\ 0.3 & 1.2\end{bmatrix}$\\
            \bottomrule
        \end{tabular}
        \caption{Parameters of the source mixture $\mu_0=\sum_{i} p_i\,\mathcal N_2(\eta_i,\Sigma_i)$ in Example~\ref{ex:gaussian_mixture_refinement}.}
        \label{tab:ex2-mu0}
    \end{table}
    
    \begin{table}[h]
        \centering
        \renewcommand{\arraystretch}{1.15}
        \setlength{\tabcolsep}{6pt}
        \begin{tabular}{cc c c c c c}
            \toprule
            $i$ & $j$ & $q_j$ & $s_{ij}^\top$ & $\sigma_{0,ij}$ & $\sigma_{1,ij}$ & $\rho_{ij}$\\
            \midrule
            1&1&0.2&$[1.8,\,0.4]$&1.0&0.7&0.5\\
            1&2&0.5&$[-0.7,\,1.6]$&0.6&1.1&-0.3\\
            1&3&0.3&$-(q_1 s_{i1}+q_2 s_{i2})/q_3$&1.2&0.5&0.1\\
            \midrule
            2&1&0.2&$[1.2,\,-1.5]$&0.9&1.2&0.7\\
            2&2&0.5&$[2.0,\,0.3]$&1.3&0.6&0.2\\
            2&3&0.3&$-(q_1 s_{i1}+q_2 s_{i2})/q_3$&0.7&1.0&-0.4\\
            \midrule
            3&1&0.2&$[-1.0,\,1.0]$&1.1&0.8&-0.6\\
            3&2&0.5&$[0.5,\,2.0]$&0.8&1.4&0.4\\
            3&3&0.3&$-(q_1 s_{i1}+q_2 s_{i2})/q_3$&1.5&0.9&0.3\\
            \bottomrule
        \end{tabular}
        \caption{Parameters of the refined target mixture $\mu_1=\sum_{i,j} p_i q_j\,\mathcal N_2(\eta_i+s_{ij},\,\Sigma_i+C_{ij})$ in Example~\ref{ex:gaussian_mixture_refinement}.}
        \label{tab:ex2-mu1}
    \end{table}
\end{example}

\begin{example}[Moons to 8 Gaussians] \label{ex:moons_to_8_gaussians}
    We conclude with a standard two-dimensional benchmark. The corresponding diagnostic plots are shown in \cref{fig:ex3-diag}. The source measure $\mu_0$ is the two-moons distribution with additive Gaussian perturbation of standard deviation $\sigma=0.1$, translated by $-[0.5,\,0.25]$ so that its mean is $[0.0,0.0]$. The target measure $\mu_1$ is the 8-Gaussians distribution obtained by placing eight Gaussian components on a circle of radius $1.75$, each with isotropic covariance $\sigma^2 I_2$ with $\sigma=0.3$.
\end{example}

\appendix

\section{Auxiliary Results} \label{sec:appendix:auxiliary_results}

\begin{theo}[Strong duality for the martingale Brenier--Benamou problem] \label{app:thm:strong_duality_mBB}
    Let $\mu_{0},\mu_{1}\in\mathcal P_{p}(\mathbb R^{d})$ for some $p > 1$ be in convex order and irreducible. Then,
    \begin{align*}
        \sup_{\pi \in \Cpl_M(\mu_0,\mu_1)} \int \MCov(\pi_x,\gamma)\, \mu_0(\rmd x) &= \inf_{\substack{\psi \in L^1(\mu_1),\\ \psi \text{ convex}}}  \left( \int \psi \,\mathrm{d}\mu_1 - \int \psi^C \,\mathrm{d}\mu_0 \right),
    \end{align*}
    where $\psi^C=(\psi^\star *\gamma)^\star$, and both sides are finite.
\end{theo}

\begin{proof}
    For every $\rho \in \mathcal{P}_{p}(\Rd)$, Young’s inequality gives
    \begin{equation}\label{ineq:bound_weak_MCov}
      \MCov(\rho,\gamma)
      \le \frac{1}{p}\!\int|y|^{p}\,\rho(\mathrm{d}y)
        + \frac{p-1}{p}\!\int|z|^{\frac{p}{p-1}}\,\gamma(\mathrm{d}z),
    \end{equation}
    and, in particular, $\int \MCov(\pi_x,\gamma)\,\mu_0(\mathrm{d}x)<\infty$ for all $\pi\in\Cpl_M(\mu_0,\mu_1)$.
    
    In analogy to \cite{backhoffveraguas2025existence}, consider the cost
    \[
        C(x,\rho) :=\begin{cases}
            -\,\MCov(\rho,\gamma) +\frac{1}{p}\!\int|y|^{p}\,\rho(\mathrm{d}y),& \bar\rho=x,\\
            \infty, & \text{otherwise},
        \end{cases}
    \]
    on $\mathbb{R}^d\times\mathcal{P}_{p}(\mathbb{R}^d)$. If we endow $\mathcal{P}_{p}$ with the corresponding Wasserstein topology, $C$ is jointly l.s.c.~and convex in $\rho$. Moreover, define
    \begin{gather*}
        \tilde\psi^C(x) := \inf_{\substack{\rho \in \mathcal{P}_{p}(\mathbb{R}^d),\\ \bar\rho = x}} \left(\int\!\psi \, \mathrm{d}\rho - \MCov(\rho,\gamma)\right), \\
        \Psi_{b,p} := \left\{\psi\colon \mathbb{R}^d\to\mathbb{R} \text{ continuous s.t. } \exists\,a,b,\ell \in \mathbb{R}\colon \ell + \tfrac{1}{p}|\cdot|^{p} \leq |\psi| \leq a+b|\cdot|^{p}\right\}.
    \end{gather*}
    Note that, choosing $\rho = \delta_x$, we have $\tilde\psi^C(x) \leq \psi(x) < \infty$. Combining \cite[Theorem 1.3]{backhoff2019existence} with \cite[Proposition 3.5]{backhoffveraguas2025existence} adapted to the growth class $\Psi_{b,p}$ yields
    \begin{equation}\label{eq:strong_duality_fixed_growth}
      \sup_{\pi\in\Cpl_M(\mu_0,\mu_1)}
          \!\int \MCov(\pi_x,\gamma)\,\mu_0(\mathrm{d}x)
      =
      \inf_{\substack{\psi\in\Psi_{b,p}\\ \psi\ \mathrm{convex}}}
         \Big( \int\psi\,\mathrm{d}\mu_1 - \int\tilde\psi^C\,\mathrm{d}\mu_0 \Big).
    \end{equation}
    
    By \cite[Proposition 5.6]{beiglbock2025fundamental}, $(\tilde\psi^C)^{\star\star} = (\psi^\star * \gamma)^\star$. To identify $\tilde\psi^C$ with its biconjugate, note that the growth condition defining $\Psi_{b,p}$ and \eqref{ineq:bound_weak_MCov} imply
    \[
      \int\psi\,\mathrm{d}\rho - \MCov(\rho,\gamma)
      \ge \ell - \frac{p-1}{p}
             \int|z|^{\frac{p}{p-1}}\,\gamma(\mathrm{d}z),
    \]
    for all $\rho\in\mathcal{P}_{p}$. Hence $\tilde\psi^C$ is finitely valued, and the Fenchel--Moreau theorem yields the desired identification:
    \[
      \tilde\psi^C = (\tilde\psi^C)^{\star\star} = (\psi^\star*\gamma)^\star=\psi^C.
    \]
    
    For any convex $\psi\in L^{1}(\mu_1)$, irreducibility and convex order of $(\mu_0,\mu_1)$ imply $\psi^C\in L^{1}(\mu_0)$ (see the proof of \cref{lem:obj_props}). By the definition of $\psi^C$, for all $\pi\in\Cpl_M(\mu_0,\mu_1)$ and $\mu_0$-a.e.\ $x$,
    \[
      \MCov(\pi_x,\gamma)
      \le \int\psi\,\mathrm{d}\pi_x - \psi^C(x),
    \]
    thus,
    \[
      \sup_{\pi\in\Cpl_M(\mu_0,\mu_1)}
         \!\int\MCov(\pi_x,\gamma)\,\mu_0(\mathrm{d}x)
      \le
      \inf_{\substack{\psi\in L^{1}(\mu_1)\\ \psi\ \mathrm{convex}}}
         \Big(\int\psi\,\mathrm{d}\mu_1-\int\psi^C\,\mathrm{d}\mu_0\Big).
    \]
    Combining this with \eqref{eq:strong_duality_fixed_growth} completes the proof.
\end{proof}
\bigskip

\begin{proof}[Proof of \cref{prop:existence_super-spreader}]
    We begin by constructing $\pi' \in \Cpl_M(\mu_0,\mu_1)$ with the claimed properties. Since the Euclidean topology on $\R^d$ admits a countable base, let $\{U_n\}_{n\in\N}$ be an enumeration of all nonempty open sets in the base with $\mu_1(U_n)>0$. It suffices to find $\pi'$ such that
    \[
      \mu_0\bigl(\{x : \pi'_x(U_n)>0 \text{ for all } n\in\N\}\bigr)=1.
    \]
    Fix $n\in\N$ and consider
    \begin{equation} \label{problem:optimal_martingale_coupling}
        \sup_{\pi \in \Cpl_M(\mu_0,\mu_1)} \mu_0\bigl(x\colon \pi_x(U_n)>0\bigr),
    \end{equation}
    which is positive by irreducibility. For any maximising sequence $(\pi_k)_{k\in\N}$, the martingale coupling 
    \[
        \pi^{(n)} := \sum_{k\in\N} 2^{-k} \pi_k
    \]
    attains \eqref{problem:optimal_martingale_coupling}. Moreover, it satisfies $\mu_0\bigl(x\colon \pi^{(n)}_x(U_n)=0\bigr) = 0$, for if this set had positive $\mu_0$-measure, we could find $\rho \in \Cpl_M(\mu_0,\mu_1)$ with ${\rho\bigl(\{x\colon \pi^{(n)}_x(U_n)=0\}\times U_n\bigr)>0}$. In this case, $\tilde \pi := \frac{1}{2}(\pi^{(n)}+\rho)$ would satisfy
    \[
        \mu_0\bigl(x\colon \pi^{(n)}_x(U_1)>0\bigr) < \mu_0(x\colon \tilde \pi_x(U_1)>0),
    \]
    a contradiction. Applying this construction to each $\{U_n\}_{n \in \N}$ yields martingale couplings $\{\pi^{(n)}\}_{n \in \N}$ such that $\pi:=\sum_{n \in \N}2^{-n} \pi^{(n)}$ satisfies the desired property.

    For the second claim, let $v$ be a Bass potential and let 
    \[
        \tilde\pi := (\operatorname{proj}_x,T)_\#(\mu_0 \otimes \gamma) \in \Cpl_M(\mu_0,\mu_1)
    \]
    with $T(x,z) := \nabla v^\star\bigl(z + \nabla v^C(x)\bigr)$ be the associated Bass martingale coupling. Let $U \subset \R^d$ be open with $\mu_1(U) > 0$. Then there exists $x_0$ with $\tilde\pi_{x_0}(U) > 0$, i.e., $\mathbb{P}[T(x_0,Z) \in U] > 0$ with $Z \sim\gamma$. If $\operatorname{supp}(\mu_0)$ is a singleton, the claim is immediate. Otherwise, writing $A := (\nabla v^\star)^{-1}(U)$ we have $\mathbb{P}\!\left[Z + \nabla v^C(x_0) \in A\right] > 0$. Since $\gamma$ has a strictly positive density, this implies $|A| > 0$ and hence $\mathbb{P}[Z + y \in A] > 0$ for all $y \in \R^d$. In particular, ${\tilde\pi_x(U) = \mathbb{P}[T(x,Z) \in U] > 0}$ for every $x$ such that $\nabla v^C(x)$ is defined. As $\operatorname{supp}(\mu_0)$ is not a singleton, $v^C$ is differentiable on an open set containing a $\mu_0$-full set by \cref{lem:ext_obj_props} and its proof, and therefore $\tilde\pi_x(U) > 0$ for $\mu_0$-a.e.~$x$.
    
    Pertaining to the final part of the proposition, we have that $\bar\mu_0 \in I_{\mu_1}$. Therefore, there exist distinct ${x_1,\dots,x_{m} \in I_{\mu_1}}$ and $\delta \in (0,1)$ so that 
    \[
        \bar B_\delta(\bar\mu_0) \subset \interior(\co(x_1,\dots,x_m)).
    \]
    We also note that $m \geq d+1$, since $\interior(\co(x_1,\dots,x_m)) = \emptyset$ otherwise. By Carath\'eodory's theorem, each vertex $x_i \in I_{\mu_1}$ can be expressed as a convex combination of at most $d+1$ points in $\supp \mu_1$, and so, without loss of generality, we assume $x_1, \dots, x_m \in \supp \mu_1$. Finally, we can choose $\delta$ still smaller such that
    \begin{align*}
        & \overline{B}_\delta(\bar\mu_0)\subset \interior(\co\{y_1,\dots,y_m\})
        &&
        \forall\, y_i\in B_\delta(x_i),\ \forall \,i=1,\dots,m, \\
        & \mu_1\bigl(B_\delta(x_i)\bigr)>0 
        &&
        \text{for $\mu_0$-a.e.\ }x,\ \forall\, i=1,\dots,m, \\
        & B_\delta(x_i)\cap B_\delta(x_j)=\emptyset
        &&
        \forall\, i\neq j. \qedhere
    \end{align*}
\end{proof}
\bigskip

\begin{lemme} \label{lem:limit_potential}
    Let $(v_n)_{n\in\N},v$ be proper l.s.c.~convex functions with $v_n \to v$ in epi-convergence and let $(\alpha_n)_{n \in \N}$, $(\mu_1^{(n)})_{n \in \N}$ be tight sequences of sub-probability measures such that $(\nabla v_n^\ast)_\# \alpha_n \ast \gamma = \mu_1^{(n)}$ for all $n \in \N$. Then, for any accumulation point $(\alpha,\mu_1)$,
    \[
        (\nabla v^\ast)_\# \alpha \ast \gamma = \mu_1.
    \]
    In particular, $\alpha_n\to \alpha$ and $\mu_1^{(n)} \to \mu_1$ weakly.
\end{lemme}

\begin{proof}
    By tightness, pass to subsequences $\alpha_n\to \alpha$ and $\mu_1^{(n)} \to \mu_1$ weakly.
    Moreover, $v_n^\ast \to v^\ast$ in epi-convergence and since $v^\ast$ is proper, there exists a sequence $\tilde z_n \to \tilde z$ with $v_n^\ast(\tilde z_n) \to v^\ast(\tilde z)$. We claim $\limsup_{n \in \N}v_n^\ast(z)<\infty$ for all $z \in \R^d$. Assuming the contrary, let $z_0 \in \R^d$ with $v_n^\ast(z_0) \to \infty$ and, potentially passing to another subsequence, choose $\hat y_n \in \partial v_n^\ast(z_0)$ with $\frac{\hat y_n}{|\hat y_n|} \to s \in S^{d-1}$. Since $v_n^\ast(z_0) \le v_n^\ast(z)$ for all $z \in \mathbb{R}^d$ with $\hat y_n \cdot (z - z_0) \ge 0$, it follows that $v_n^\ast \to \infty$ on $H^+_{s,0}(z_0) := \{ z \in \R^d : s \cdot  (z - z_0) > 0 \}$.
    For every $(z,y_n) \in \partial v_n^\ast \cap H^+_{s,0}(z_0) \times \R^d$,
    \[
        \frac{v_n^\ast(z) - v_n^\ast(\tilde z_n)}{|z - \tilde z_n|} \le
        y_n \cdot \frac{(z - \tilde z_n)}{|z - \tilde z_n|} \le |y_n|,
    \]
    hence $|y_n| \to \infty$. Moreover, tightness of $(\alpha_n)_{n \in \N}$ yields $\inf_{n \in \N} \alpha_n \ast \gamma\!\left(H^+_{s,0}(z_0)\right) \ge \delta$ for some $\delta > 0$.
    Therefore, for any $R>0$,
    \[
        \lim_{n \to \infty} \alpha_n\ast\gamma(|\nabla v_n^\ast| \le R) \le 1 - \delta,      
    \]
    so $(\mu_1^{(n)})_{n \in \N}$ is not tight, a contradiction. Consequently, $v^{\FL}$ is finite everywhere.
    
    Since $v^\ast$ is real-valued, $v_n^\ast \to v^\ast$ pointwise and so we have $\nabla v_n^\ast \to \nabla v^\ast$ almost everywhere (see \cref{lem:properties_epi_convergence} and subsequent remarks). Truncating the gradients and applying Egorov's theorem, we conclude 
    \[
        (\nabla v_n^\ast)_\# \alpha_n \ast \gamma \to (\nabla v^\ast)_\# \alpha \ast \gamma \ \text{ weakly}. \qedhere
    \]
\end{proof}
\bigskip

\begin{lemme} \label{lem:dense_unique_subdifferential}
    Let $f : \R^d \to \R$ be convex, and let $D \subseteq \partial f$ be such that $\{x : (x,y) \in D\}$ is dense in $\R^d$.
    Then $D$ determines $f$ uniquely up to an additive constant.
\end{lemme}

\begin{proof}
    Let $g : \R^d \to \R$ be another convex function with $D \subseteq \partial g$.
    Since $f$ and $g$ are continuous and convex, $\nabla f$ and $\nabla g$ exist almost everywhere by Rademacher's theorem.
    Let $x \in \R^d$ for which $\nabla f(x)$ and $\nabla g(x)$ exist and pick $(x_n,y_n)_{n \in \N}$ in $D$ such that $x_n \to x$. By \cite[Theorem 24.5]{R} we have $y_n \to \nabla f(x)$ and $y_n \to \nabla g(x)$.
    Thus, $\nabla f = \nabla g$ almost everywhere.
    By the Poincar\'e--Wirtinger inequality, for every $R > 0$,
    \[
        \int_{B_R(0)} |f(x) - g(x) - m_{f,R} - m_{g,R}| \, \mathrm{d}x \le C \int_{B_R(0)} |\nabla f(x) - \nabla g(x)| \, \mathrm{d}x = 0,
    \]
    for some constant $C > 0$, with
    \[
        m_{f,R} := \int_{B_R(0)} f(x) \, \frac{\mathrm{d}x}{{\rm Vol}(B_R(0))}, \qquad
        m_{g,R} := \int_{B_R(0)} g(x) \, \frac{\mathrm{d}x}{{\rm Vol}(B_R(0))}.
    \]
    Hence, $f - m_{f,R} - m_{g,R} = g$ on $B_R(0)$, for every $R > 0$.
\end{proof}
\bigskip

\begin{lemme} \label{lem:disjoint_subdiffs}
    Let $f,g : \R^d \to \R$ be convex functions such that $f - g$ is not constant.
    Then, there exists an open $B \subseteq \R^d$ with $\partial f(x) \cap \partial g(x) = \emptyset$ for all $x \in B$.
\end{lemme}

\begin{proof}
    Assume the contrary, i.e., for every open set $B$ we can find $(x,y) \in B \times \R^d$ with ${y \in \partial f(x) \cap \partial g(x)}$.
    Let $(B_n)_{n \in \N}$ be a countable base of the topology on $\R^d$, and denote by $(x_n,y_n) \in B_n \times \R^d$ a pair with $y_n \in \partial f(x_n) \cap \partial g(x_n)$.
    Set $D := \{(x_n,y_n) : n \in \N\}$. Then, $\{x_n : n \in \N\}$ is dense in $\R^d$, so $f - g$ must be constant by \cref{lem:dense_unique_subdifferential}, a contradiction.
\end{proof}
\bigskip

\begin{lemme} \label{lem:exp.bound}
    Let $(f_n)_{n \in \N}$ be real-valued convex functions such that $\sup_{n\in\N} |f_n| < \infty$ on $B_\delta(0)$ for some $\delta > 0$. If, for some $p > 1$,
    \begin{align*}
        \int |\nabla f_n|^{p} \, \mathrm{d}\gamma < \infty,
    \end{align*}
    then there exists $c, R > 0$ such that, for all $|x| \ge R$,
    \[
        \sup_{n \in \N} |f_n(x)| \le c\exp\!\left(\tfrac2{3+p}|x|^2 \right).
    \]
\end{lemme}

\begin{proof}
    We assume $\delta = 1$ for simplicity. Convexity and the uniform bound on $B_1(0)$ yield $\inf_{n \in \N} |f_n(x)| > -\infty$ for all $x\in\Rd$.
    Without loss of generality we may assume $-1 \le f_n \le 0$ on $B_1(0)$ for all $n\in\mathbb N$, after an affine rescaling of the $(f_n)_{n \in \N}$. This only affects the multiplicative constant $c$ in the estimates below. Thus it remains to show that there exists $R > 0$ such that, for all $|x| \ge R$,
    \[
        \sup_{n \in \N} f_n(x) \le \exp\!\left(\tfrac2{3+p}|x|^2 \right).
    \]
    Assume the contrary, i.e., potentially passing to a subsequence, there is a sequence $(x_n)_{n \in \N}$ with $|x_n| \to \infty$, $|x_n| \ge 1$, and $f_n(x_n) \ge \exp\!\left(\tfrac2{3+p}|x_n|^2 \right)$.
    Let $n \in \N$. Since $f_n$ is convex, we have for a.e.~$z \in B_1(0)$ and $t \geq 0$
    \begin{align*}
        \frac{f_n(x_n) - f_n(z)}{|x_n - z|} \le
        \frac{f_n(x_n + t(x_n-z))-f_n(z + t(x_n-z))}{|x_n-z|}
        \le
        \nabla f_n(x_n + t(x_n-z)) \cdot \frac{(x_n-z)}{|x_n-z|},
    \end{align*}
    which yields
    \[
        \frac{\exp\!\left(\tfrac2{3+p}|x_n|^2 \right)}{|x_n| + 1} \le \frac{f_n(x_n) - f_n(z)}{|x_n-z|} \le |\nabla f_n(x_n + t(x_n-z))|. 
    \]
    With $C_n := \{ x_n + t(x_n - z) : z \in B_1, t \ge 0 \}$ and by \cref{lem:cone_lower_bound}, for some $\varepsilon > 0$,
    \[
        \int |\nabla f_n|^{p} \, \mathrm{d}\gamma \ge
        \frac{\exp\!\left(\frac{2p}{3+p} |x_n|^2\right)}{(|x_n|+1)^p} \gamma(C_n)
        \ge c \frac{\exp(\varepsilon |x_n|^2 - |x_n|)}{(|x_n| + 1)^p},
    \]
    Passing to the limit, we have $\int |\nabla f_n|^{p} \, \mathrm{d}\gamma \to \infty$, a contradiction.
\end{proof}

\begin{lemme} \label{lem:cone_lower_bound}
    Let $x \in \R^d$ and $C = \{ x + t(x-z) : z \in B_1(0), t\ge 0\}$.
    Then, for all $p > 1$ there exist constants $c = c(d) > 0$ and $\varepsilon >0$ with
    \[
        \gamma(C) \ge c \exp\!\left(-\left(\tfrac{2p}{3+p}-\varepsilon\right)|x|^2 - |x|\right).
    \]
\end{lemme}

\begin{proof}
    Since $x + t (x - z) = (1 + t) x - t z \in C$, we have $B_t\bigl((1+t)x\bigr) \subseteq C$. Therefore,
    \[
        \gamma(C) \ge \gamma\!\left(B_t\bigl((1+t)x\bigr)\right) \ge \frac{{\rm Vol}(B_t)}{(2\pi)^{d/2}} \exp\!\left(-\tfrac{1}{2}((1+t)|x| + t)^2\right)
    \]
    With $t = \sqrt{2}\sqrt{\frac{p}{1+p}} -1 \geq 0$, we have $\frac{1}{2}(1+t)^2=\frac{p}{1+p}$ and hence constants $c = c(d) > 0$ and $\varepsilon > 0$ such that
    \[
        \gamma(C) \ge c \exp\!\left(-\left(\tfrac{2p}{3+p}-\varepsilon\right)|x|^2 - |x|\right). \qedhere
    \]
\end{proof}
\bigskip

\begin{lemme} \label{app_B:lem:existence_c1_convex_coeff}
    Let $A \subset \R^d$ be an affine subspace and let $U \subset A$ be non-empty, convex and open, and let $f \in \mathcal{C}^1(U,U)$, all with respect to the relative topology on $A$. Define
    \[
        \mathcal{H}^\circ_m(U):=\left\{(v_1,\dots,v_m,x) \in U^{m+1}\colon \ f(x) \in \operatorname{int}_A(\co(v_1,\dots,v_m))\right\},
    \]
    where $\operatorname{int}_A(B)$ denotes the interior of $B \subset A$ in the relative topology on $A$.
    Then there exists $\alpha \in \mathcal{C}^1\bigl(\mathcal{H}^\circ_m(U),(0,1)^m\bigr)$ such that, for all $(v_1,\dots,v_m,x) \in \mathcal{H}^\circ_m(U)$,
    \[
        f(x) = \sum_{i=1}^m \alpha_i(v_1,\dots,v_m,x)v_i.
    \]
\end{lemme}

\begin{remarque}
    Let $k = \dim(A)$. Then $\mathcal{H}^\circ_m(U)$ is non-empty if and only if $U \neq \emptyset$ and $m \geq k+1$. Moreover, a standard separating hyperplane argument shows that $\mathcal{H}^\circ_m(U)$ is open in $A^{m+1}$ for the relative topology.
\end{remarque}

\begin{proof}
    By working in affine coordinates on $A$, we may assume without loss of generality that $A = \R^d$. Further assume $U\neq \emptyset$ and $m \geq d+1$, since otherwise there remains nothing to show. Fix $v_1, \dots, v_m \in U$ and let $V := [v_1,\dots,v_m] \in \R^{d \times m}$ be the matrix whose columns are the points spanning the convex polytope $\co(v_1,\dots,v_m)$. For $x\in\interior(\co(v_1,\dots,v_m))$, consider the convex optimization problem
    \begin{align*}
        \inf_{\alpha \in (0,1)^m} \sum_{i=1}^m \alpha_i\log(\alpha_i) \quad \text{s.t.} \quad \begin{cases}
            V\alpha = f(x), \\
            \displaystyle \sum_{i=1}^m \alpha_i = 1.
        \end{cases}
    \end{align*}
    By construction, the feasible set is non-empty, so the infimum is attained in $(0,1)^m$. Moreover, the Lagrangian is
    \[
        \mathcal{L}(\alpha,\theta,\kappa) = \sum_{i=1}^m \alpha_i\log(\alpha_i) - \langle \theta, V\alpha - f(x)\rangle - \kappa \left(\sum_{i=1}^m \alpha_i - 1\right),
    \]
    where $\alpha \in (0,1)^m$, $\theta \in \R^d$, and $\kappa \in \R$, and its stationary point is given implicitly by
    \begin{align*}
        \kappa(\theta) &= \sum_{i=1}^m e^{-\langle\theta,v_i\rangle}, & \alpha_i(\theta) &= \frac{e^{-\langle\theta,v_i\rangle}}{\sum_{i=1}^m e^{-\langle\theta,v_i\rangle}} \quad (i \in \{1, \dots, m\}), & f(x) &= \sum_{i=1}^m v_i \alpha_i(\theta).
    \end{align*}
    The implicit function theorem yields $\theta \in \mathcal{C}^1\bigl(\mathcal{H}^\circ_m(U),\R^d\bigr)$ so that, for all ${(v_1,\dots,v_m,x) \in \mathcal{H}^\circ_m(U)}$,
    \[
        f(x) = \sum_{i=1}^m v_i \alpha_i\bigl(v_1,\dots,v_m,\theta(v_1,\dots,v_m,x)\bigr). \qedhere
    \]
\end{proof}

\printbibliography

\clearpage
\thispagestyle{empty}
\begin{figure}[t]
    \centering
    
    \begin{subfigure}[b]{0.55\textwidth}
        \centering
        \includegraphics[width=\textwidth]{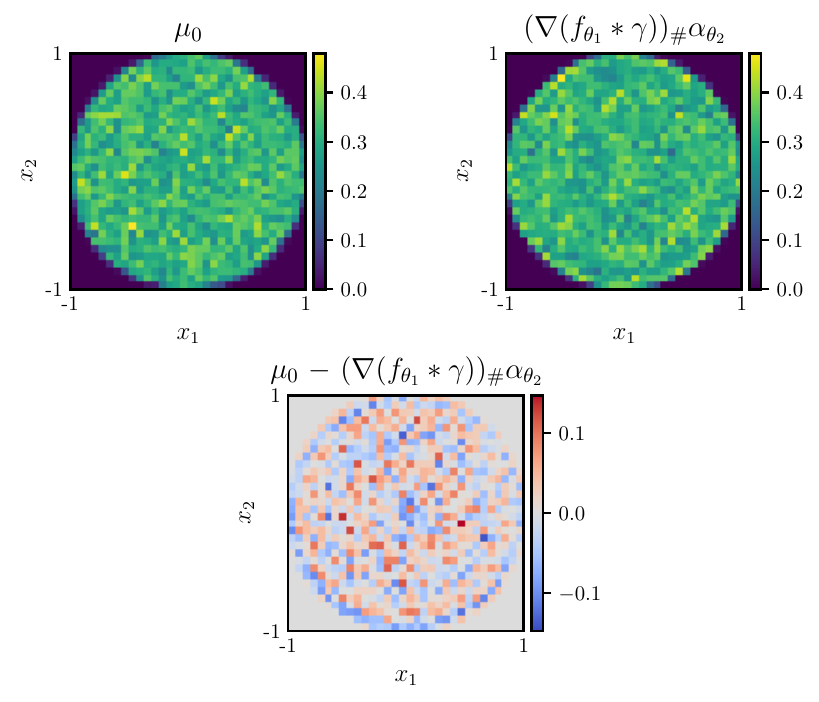}
    \end{subfigure} \hfill
    
    \begin{subfigure}[b]{0.55\textwidth}
        \centering
        \includegraphics[width=\textwidth]{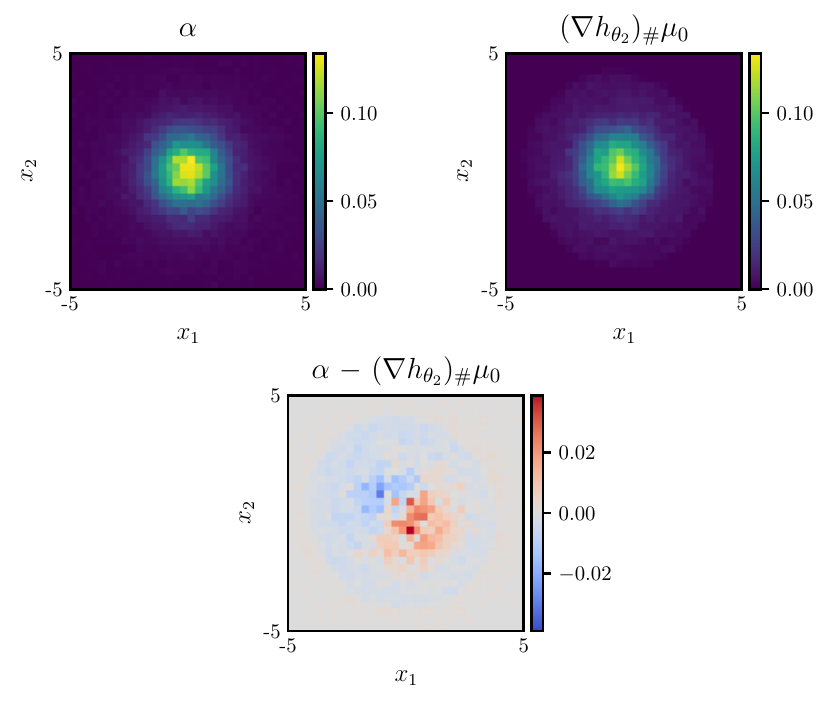}
    \end{subfigure} \hfill

    \begin{subfigure}[b]{0.55\textwidth}
        \centering
        \includegraphics[width=\textwidth]{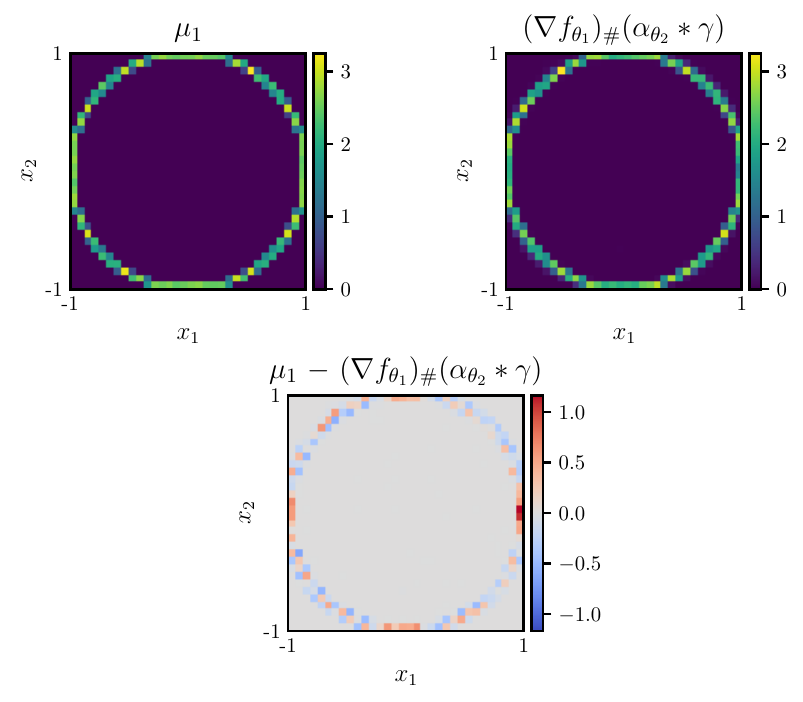}
    \end{subfigure}
    
    \caption{Diagnostic plots for \cref{ex:uniform_disk_to_circle}. From top to bottom:
    (i) $\mu_0$ and its empirical reconstruction $(\nabla(f_{\theta_1}*\gamma))_\#\alpha_{\theta_2}$.
    (ii) $\alpha$ and its learned approximation $(\nabla h_{\theta_2})_\#\mu_0$.
    (iii) $\mu_1$ and its empirical reconstruction $(\nabla f_{\theta_1})_\#(\alpha_{\theta_2}*\gamma)$.}
    \label{fig:ex1-diag}
\end{figure}
\clearpage

\clearpage
\thispagestyle{empty}
\begin{figure}[t]
    \centering
    
    \begin{subfigure}[b]{0.55\textwidth}
        \centering
        \includegraphics[width=\textwidth]{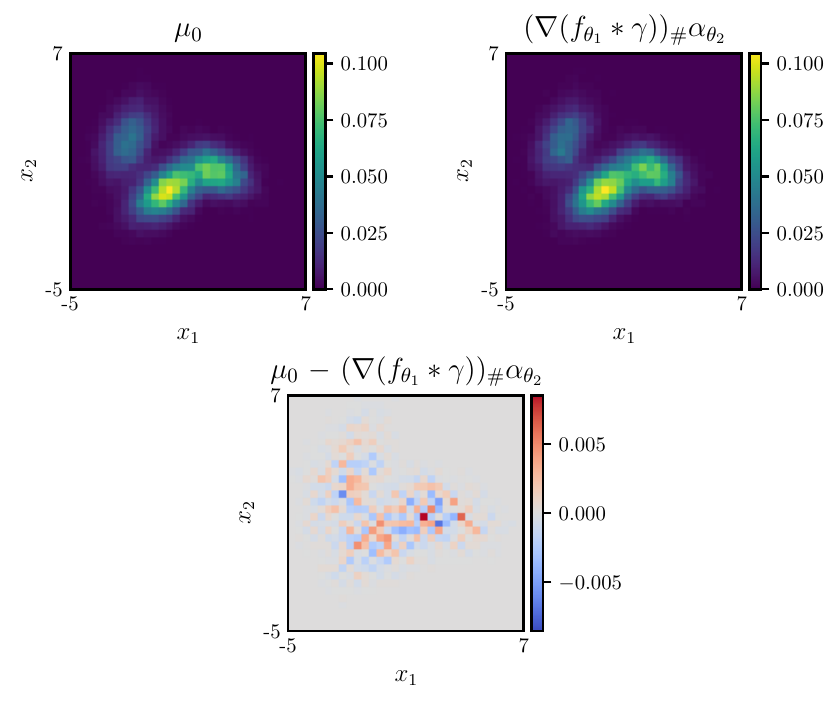}
    \end{subfigure} \hfill
    
    \begin{subfigure}[b]{0.55\textwidth}
        \centering
        \includegraphics[width=\textwidth]{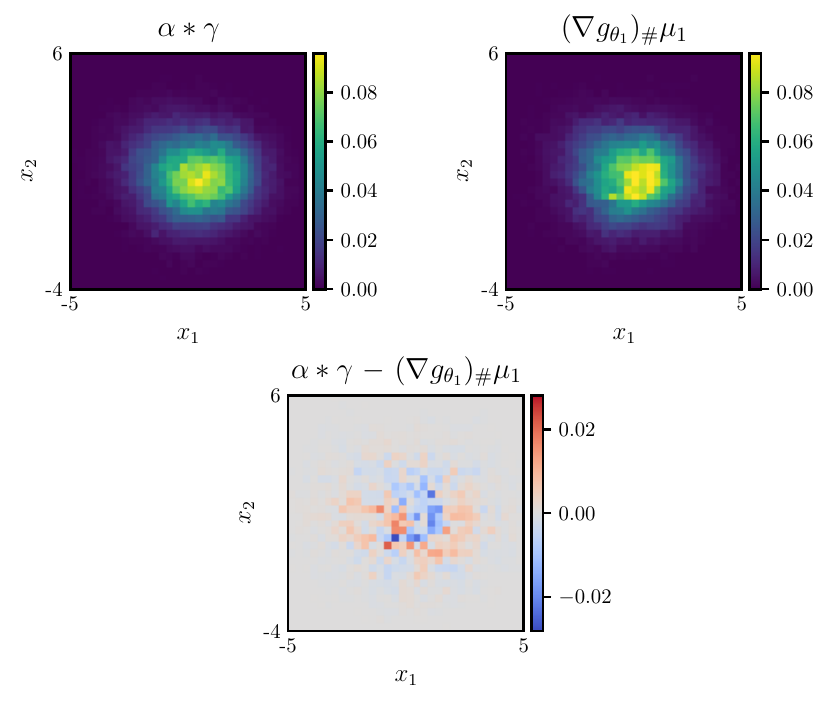}
    \end{subfigure} \hfill

    \begin{subfigure}[b]{0.55\textwidth}
        \centering
        \includegraphics[width=\textwidth]{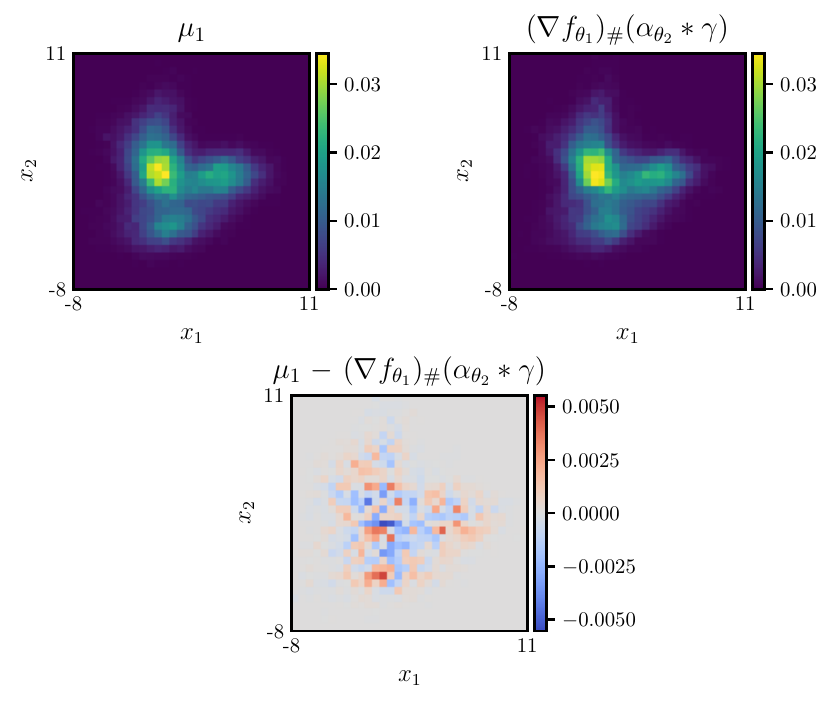}
    \end{subfigure}
    
    \caption{Diagnostic plots for \cref{ex:gaussian_mixture_refinement}. From top to bottom: 
    $\mu_0$ and its empirical reconstruction $(\nabla(f_{\theta_1}*\gamma))_\#\alpha_{\theta_2}$;
    $\alpha_{\theta_2}*\gamma$ and its learned approximation $(\nabla g_{\theta_1})_\#\mu_1$;
    $\mu_1$ and its empirical reconstruction $(\nabla f_{\theta_1})_\#(\alpha_{\theta_2}*\gamma)$;}
    \label{fig:ex2-diag}
\end{figure}
\clearpage

\clearpage
\thispagestyle{empty}
\begin{figure}[t]
    \centering
    
    \begin{subfigure}[b]{0.55\textwidth}
        \centering
        \includegraphics[width=\textwidth]{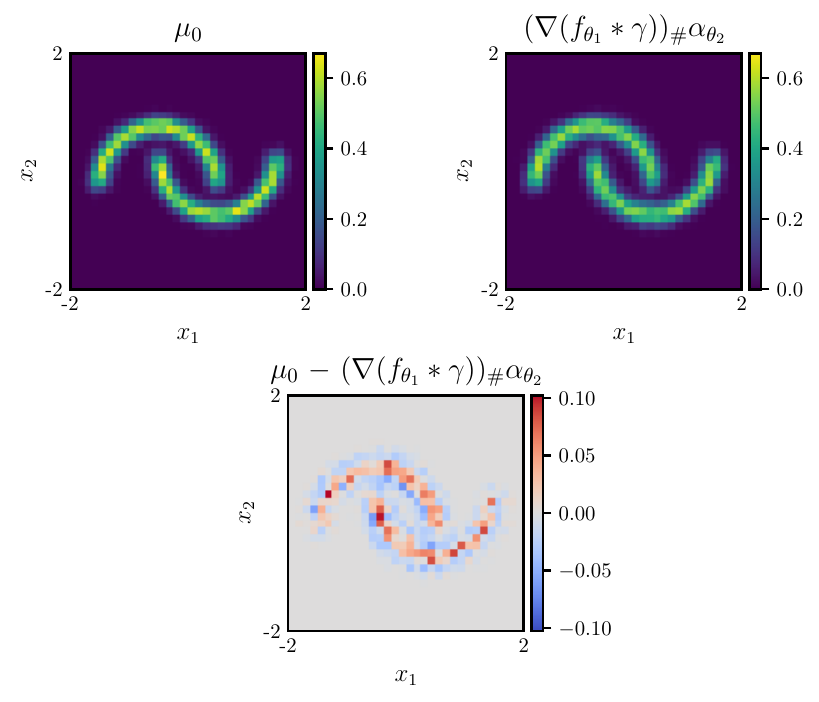}
    \end{subfigure} \hfill
    
    \begin{subfigure}[b]{0.55\textwidth}
        \centering
        \includegraphics[width=\textwidth]{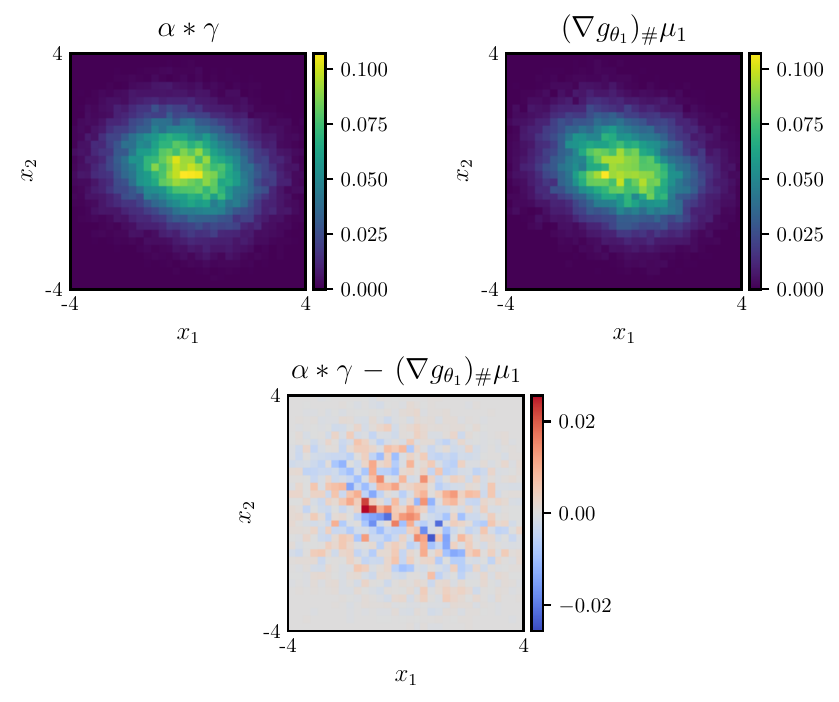}
    \end{subfigure} \hfill

    \begin{subfigure}[b]{0.55\textwidth}
        \centering
        \includegraphics[width=\textwidth]{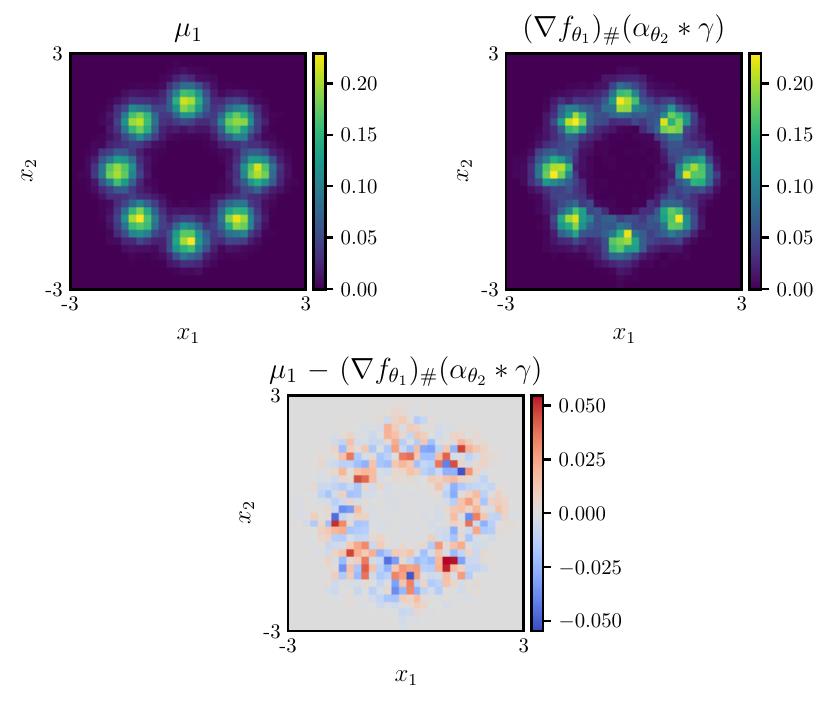}
    \end{subfigure}
    
    \caption{Diagnostic plots for \cref{ex:moons_to_8_gaussians}. From top to bottom: 
    $\mu_0$ and its empirical reconstruction $(\nabla(f_{\theta_1}*\gamma))_\#\alpha_{\theta_2}$;
    $\alpha_{\theta_2}*\gamma$ and its learned approximation $(\nabla g_{\theta_1})_\#\mu_1$;
    $\mu_1$ and its empirical reconstruction $(\nabla f_{\theta_1})_\#(\alpha_{\theta_2}*\gamma)$;}
    \label{fig:ex3-diag}
\end{figure}
\clearpage
\end{document}